\documentclass[12pt,a4paper]{article}

\usepackage{a4wide}
\usepackage{amsmath}
\usepackage{amssymb}
\usepackage{amsfonts}
\usepackage{epsfig}
\usepackage{exscale}
\usepackage{float}
\usepackage{bbm}
\usepackage[numbers,sort&compress]{natbib}

\newcommand{\R}{{\mathbb{R}}}

\makeatletter
\@addtoreset{equation}{section}
\makeatother

\title{The General Solution for the Relativistic and Nonrelativistic  Schr\"odinger Equation for the $\delta^{(n)}$-Function Potential \\in 1-dimension
Using Cutoff Regularization, and the Fate of Universality}

\author{M.\ H.\ Al-Hashimi$^{a,b}$, M.\ Salman $^{b}$,   A.\ M.\ Shalaby$^{b}$
\footnote{Contact information: M.\ H.\ Al-Hashimi: hashimi@itp.unibe.ch, M.\ Salman: msalman@qu.edu.qa, +974 4403 4618
+41 31 631 8878; A.\ Shalaby, amshalab@qu.edu.qa, +974 4403 4630.}
\\ \\
$^a$ Albert Einstein Center for Fundamental Physics \\
Institute for Theoretical Physics, University of Bern\\
Sidlerstrasse 5, CH-3012 Bern, Switzerland \\ \\
$^b$ Department of Mathematics, Statistics, and Physics \\
Qatar University, Al Tarfa, Doha 2713, Qatar
}

\begin{document}

\maketitle

\vspace{-1cm}

\begin{abstract} \normalsize
A general method has been developed to solve the Schr\"odinger  equation for an arbitrary derivative of the $\delta$-function potential in 1-d using cutoff regularization. The work treats both the relativistic and nonrelativistic cases. A distinction in the treatment has been made between the case when the derivative $n$ is an even number from the one when $n$ is an odd number. A general gap equations for each case has been derived. The case of  $\delta^{(2)}$-function potential has been used as an example. The results from the relativistic case show that the $\delta^{(2)}$-function system behaves exactly like the $\delta$-function and  the $\delta'$-function potentials, which means it also shares the same features with quantum field theories, like being asymptotically free, in the massless limit, it undergoes dimensional transmutation and it possesses an infrared conformal fixed point. As a result the evidence of universality of contact interactions has been extended further to include the $\delta^{(2)}$-function potential.
\end{abstract}
\section{Introduction}
Contact interaction has been investigated nonrelativistically in numerous studies using different methods in the context of the non-relativistic Schr\"odinger equation \cite{Cal88,Tho79,Beg85,Hag90,Jac91,Fer91,Gos91,Mea91,Man93,Phi98,Hold,Albbook,Arnbak1,Zhao,Griff,Toy,Widmer,Fas013,lange,Fas015}, and in the context of Dirac equation \cite{Cal14,Arnbak11}. Regularization is an approach that is widely used in quantum field theories \cite{Bol72,Bol72a,tHo72,Bie14}.The solution of the Schr\"odinger equation in 1-d for $\delta^{(n)}$-potential needs to be regularized when $n\geq 1$  \cite{Widmer,Phi98} even for non-relativistic solution.  The relativistic contact interaction potentials have been investigated in much smaller number of articles, in the context of self-adjoint extensions for pseudodifferential operators by using abstract mathematical approach. This approach dose not require the use of any regularization method, and the concept of a wave function is not necessary to obtain physical quantities like the scattering amplitude, or bound state \cite{Alb97}. However the mathematical language, and the treatment is beyond the grasp of most physicists.

An important feature of most of quantum field theories is locality \cite{Frank,UweNots}. In non-relativistic quantum mechanics, according to the theory of self-adjoint extensions, the kinetic part of the Hamiltonian $p^2/(2m)$ is local, which means that the wave function to the left and to the right of the contact interaction is a wave function of a free particle. The boundary condition is characterized by a family of self-adjoint extension parameters. In 1-dimension, the boundary condition for the wave function at contact point takes the following form
\begin{equation}\label{Self1D1}
\left(\begin{array}{c} \Psi(\varepsilon) \\ \partial_x \Psi(\varepsilon)
\end{array}\right) =
\exp(i \theta) \left(\begin{array}{cc} a & b \\ c & d \end{array}\right)
\left(\begin{array}{c} \Psi(-\varepsilon) \\ \partial_x \Psi(-\varepsilon)
\end{array}\right),
\end{equation}
where $\varepsilon \rightarrow 0$, $a, b, c, d \in \R$. In addition, the parameters subject to the condition $ad - bc = 1$, and
$\theta \in ]- \frac{\pi}{2},\frac{\pi}{2}]$. Therefore, the five parameters
$a, b, c, d, \theta$ are reduced to 4-parameter family of self-adjoint extensions of the non-relativistic
free-particle Hamiltonian, which can describe any contact interaction. For example the $\delta^{(n)}$-function, and after imposing parity symmetry, the boundary condition can be reduced to a 1-parameter family of self-adjoint extensions, which is
\begin{equation}\label{Self1D2}
\left(\begin{array}{c} \Psi(\varepsilon) \\ \partial_x \Psi(\varepsilon)
\end{array}\right) =
\exp(i \theta) \left(\begin{array}{cc} a & b \\ c & d \end{array}\right)
\left(\begin{array}{c} \Psi(-\varepsilon) \\ \partial_x \Psi(-\varepsilon)
\end{array}\right).
\end{equation}
where
\begin{equation}\label{EB}
    \Delta E_B=-\frac{\varkappa^2}{2m},
\end{equation}
and $\Delta E_B$ is the binding energy of the of the particle with mass $m$.

Recently, the problem has been investigated by solving directly the relativistic Schr\"odinger equation in 1-dimension for the $\delta$-function potential and the $\delta'$-function potential. The problem requires  dimensional or cutoff regularization \cite{Our,Our1}. The resulting wave function for the bound state is
\begin{equation}\label{deltaWF}
\Psi_B(x) = \lambda \Psi_B(0) \left[\frac{1}{\pi}  \int_m^\infty d\mu
\frac{\sqrt{\mu^2 - m^2}}{E_B^2 - m^2 + \mu^2} \exp(- \mu |x|) +
\frac{E_B \exp(-\sqrt{m^2 - E_B^2} |x|)}{\sqrt{m^2 - E_B^2}}\right],
\end{equation}
where $E_B$ is the energy of the bound state. The resulting wave function for the scattering states is
\begin{eqnarray}\label{PsiXdeltaFirst}
\Psi_E(x)&=& A e^{ikx} +B e^{-ikx}+
\lambda(E,E_B)(A+B)\sqrt{k^2 + m^2} \frac{\sin(k |x|)}{k}
\nonumber \\
&-&\frac{1}{\pi}\lambda(E,E_B)(A+B)\int_m^\infty d\mu
\frac{\sqrt{\mu^2 - m^2}}{\mu^2 + k^2} \exp(- \mu |x|).
\end{eqnarray}
where $E$ is the energy of a scattering state, $\lambda(E,E_B)$ is the renormalized coupling constant, $A,B$ are constants. Aside from the contact point $(x=0)$, the same result can be obtained for $\delta'$-function potential. It is important to notice here that; the argument that the wave function to the left and to the right of the origin does not feel the contact interaction, is no longer valid. That is because of the second term in eq.(\ref{PsiXdeltaFirst}), which does not vanish for $x\neq 0$. This is one of the reasons that makes eq.(\ref{Self1D1}) or eq.(\ref{Self1D2}) invalid for the relativistic case. That is why we say that pseudodifferential $\sqrt{p^2 + m^2}$ is non-local. In fact, this is one of the most important results of \cite{Our,Our1,Our2}. The treatment also shows that the $\delta$-function potential and  $\delta'$-function potential shares several non-trivial features with
relativistic quantum field theories. For example, it is asymptotically free \cite{Gro73,Pol73}, just like quantum chromodynamics (QCD) \cite{Fri73}. In addition, in the massless limit, it undergoes
dimensional transmutation, and it possesses an infrared conformal fixed point. An additional important feature that  relativistic mechanics shares with some quantum field theories is universality. In \cite{Our}, it has been shown that we can not distinguish physically between the $\delta$-function potential, the $\delta'$-function potential, or a combination of them. This is similar to the situation in local quantum field theories, when all the Lagrangians corresponding to different models reduce to one Lagrangian once the cutoff is removed. The evidence of universality for the solution of the relativistic Schr\"odinger equation with a general contact interactions in 1-dimension is not conclusive. That is because higher derivative than one has never been examined relativistically in this context for the $\delta$-function potential.

The method of choice for regularizing the is cutoff regularization. It has many advantages, one of the most important is; it gives us a quantitative measure of how big is big and how small is small in terms of the momentum cutoff $\Lambda$. For example, the gap equation for the $\delta$-function potential case is
\begin{equation}\label{lambS1}
    \lambda(\Lambda)=\frac{1}{I(E_B,\Lambda)},
\end{equation}
 where
\begin{eqnarray}\label{lambS2}
I(E_B,\Lambda)=\frac{1}{2\pi } \int_{-\Lambda }^{\Lambda }\frac{1}{E_B-%
\sqrt{p^{2}+m^{2}}}dp
\end{eqnarray}
From eq(\ref{lambS2}) and eq(\ref{lambS1}), it is obvious that $\lambda(\Lambda)\rightarrow 1/\log\Lambda=0$ as $\Lambda \rightarrow \infty$. On the other hand  eq.(\ref{deltaWF}) gives
\begin{equation}\label{PsiSm}
  \Psi_B(0)=C_1 I(E_B),
\end{equation}
where
\begin{equation}\label{ILim}
    I(E_B)=\lim_{\Lambda\rightarrow \infty} I(E_B,\Lambda),
\end{equation}
and $C_1$ is the normalization constant. Eq.(\ref{PsiSm}) means that $\Psi_B(0)\rightarrow \infty$ as $\Lambda \rightarrow \infty$, while $\lambda_1\Psi_B(0)=C_1$ is a finite non zero quantity. In the case of the $\delta'$-function potential, $\lambda_1\Psi_B'(0)$ is finite while $\lambda_1\Psi_B(0)\rightarrow 0$ as $\Lambda\rightarrow\infty$. The momentum cutoff is not only envisage how fast functions go to infinity or to zero at certain point, it also reveal the behavior of these functions near a singular point or points. It is important to remember that, what is right for a very large $\Lambda$ is right for $\Lambda\rightarrow \infty$. In this way, we can not just avoid dealing with a function at singular points because they are undefined. To highlight this point, let us introduce the momentum cutoff to the $\delta$-function. Then we can write
\begin{equation}\label{deltaCut}
    \delta(x)=\lim_{\Lambda\rightarrow \infty}\delta(x,\Lambda)=\lim_{\Lambda\rightarrow \infty} \frac{1}{2\pi} \int_{-\Lambda }^{\Lambda } \exp(ipx)dp.
\end{equation}
It straightforward to find that the  extreme  value of  $\delta(x,\Lambda)$ at the origin is proportional to $\Lambda$. A more delegate example is the $\delta'$-function. For this case
\begin{equation}\label{deltaCut}
    \delta'(x)=\lim_{\Lambda\rightarrow \infty}\delta'(x,\Lambda)=\lim_{\Lambda\rightarrow \infty} \frac{1}{2\pi} \int_{-\Lambda }^{\Lambda }ip \exp(ipx)dp.
\end{equation}
\begin{figure}[tbh]
\begin{center}
\includegraphics[bb=280 28 400 400,scale=0.6]{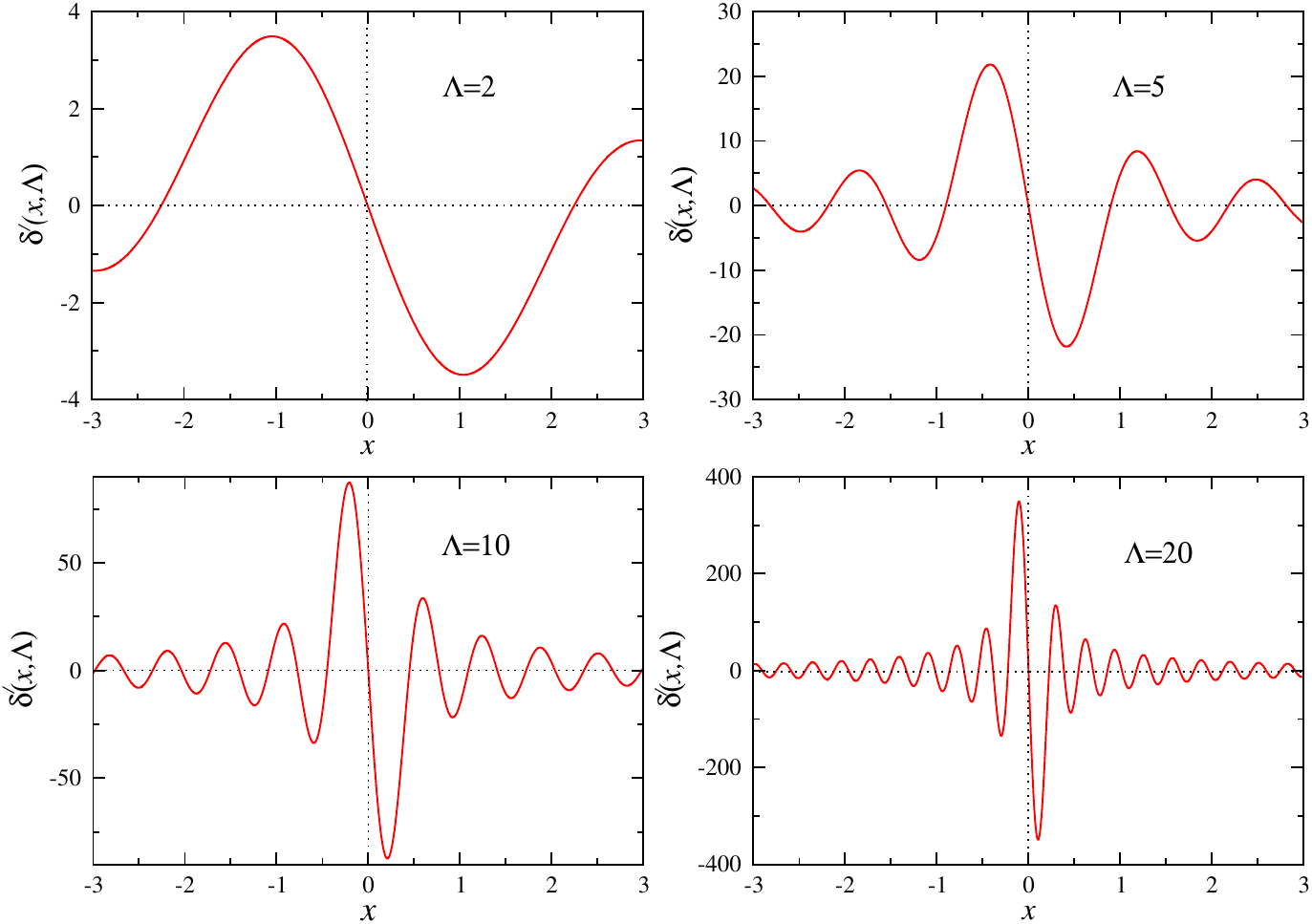}
\vspace{5mm}
\caption{\it A plot for the derivative of the dirac delta function with different values of $\Lambda =2,5,10$  and $20$.}
\label{fig1}
\end{center}
\end{figure}
Form Figure 1, and as expected from an odd function, it vanishes at origin . The nearest extrema of the function $\delta'(x,\Lambda)$ to the origin are at $x=\pm\varsigma$. The extreme values are  proportional to $\Lambda$, and the value of $ \varsigma$ is inversely proportional to $\Lambda$. The previous two examples show that the behavior of a singular function at, or near the origin, can be understood by cutting off the integral interval from $(-\infty,\infty)$ to $[-\Lambda,\Lambda]$.

The present work is aiming to present a general scheme for solving the Schr\"odinger equation with arbitrary derivative of the $\delta$-potential using cutoff regularization. Both of the non-relativistic and relativistic cases are studied in details. The $\delta^{(2)}$-potential is presented as an example for this general treatment, it has been shown that universality contact interaction holds for this case too. Before removing the cutoff, it has been proved that there are two gap equations for the even function solution, however, after removing the cutoff the number of parameters reduce to only one parameter in both of the non-relativistic and relativistic cases. In fact this is the correct number of parameters obtained from the non-relativistic theory  of self-adjoint extension.
The work has lead to an addition new analogy between relativistic quantum mechanics and quantum field theories; it was proved that relativistic case leads to a solution that is reduced to the trivial free particle solution once the cutoff is removed.

\section{The Non-Relativistic Solution}
The solution of non-relativistic $\delta ^{(n)}(x)$ -function potential problem can be studied using certain procedure of cutoff regularization.  This provides an important guidance of how to approach the relativistic case. The non-relativistic Schr\"odinger equation in this case is
\begin{equation}\label{Schrodinger3}
\frac{p^{2}}{2m}\Psi (x)+\lambda_n \delta ^{(n)}(x)\Psi (x)=\Delta E\Psi (x),
\end{equation}
where $\lambda_n$ is the bare coupling constant. In momentum space, the above equation is
\begin{equation}\label{pSpaceNon1}
\frac{p^{2}}{2m}\widetilde{\Psi }(p)+\lambda_n \int_{-\infty }^{\infty
}\delta ^{(n)}(x)\Psi (x)e^{-ipx}dx=\Delta E\widetilde{\Psi }(p),
\end{equation}
where
\begin{eqnarray}\label{FourierT}
\Psi^{(k)}(x) = \frac{1}{2 \pi} \int dp \ (ip)^{k} \widetilde \Psi(p) \exp(i p x), \quad
\Psi^{(k)}(0) = \frac{1}{2 \pi} \int dp \ (ip)^{k} \widetilde \Psi(p).
\end{eqnarray}
The second term in eq.(\ref{pSpaceNon1}) can be written as
\begin{equation}\label{PatialInt}
\int_{-\infty }^{\infty }\delta ^{(n)}(x)\Psi
(x)e^{-ipx}dx=-\int_{-\infty }^{\infty }\delta^{(n-1)} (x)\frac{d\Psi (x)}{dx}%
e^{-ipx}dx+ip\int_{-\infty }^{\infty }\delta^{(n-1)} (x)\Psi (x)e^{-ipx}dx.
\end{equation}
The partial integral can be repeated in the above equation. From the general Leibniz rule, we have
\begin{equation}
\frac{d^{n}}{dx^{n}}(f(x)g(x))=\sum_{j=0}^{n}C^{n}_{j}
 \frac{d^{n-j}}{dx^{n-j}}f(x)\frac{d^{j}}{dx^{j}}g(x),
\end{equation}

where $C^{n}_{j}$ is the binomial coefficient. In addition, is easy to prove that
\begin{equation}
\frac{d^{j}}{dx^{j}}e^{-ipx}|_{x=0}=(-ip)^{j},
\end{equation}
therefore eq.(\ref{PatialInt}) can be written as
\begin{eqnarray}\label{FnpEq}
\int_{-\infty }^{\infty }\delta ^{(n)}(x)\psi (x)e^{-ipx}dx=\frac{d^{n}}{%
dx^{n}}(\Psi (x)e^{-ipx})|_{x=0}\nonumber\\=\sum_{j=0}^{n}
C^{n}_{j} \Psi ^{(n-j)}(0)(-ip)^{j}\equiv F(n,p)
\end{eqnarray}

\begin{equation}\label{GeneralPsiNon}
\Psi (x)=\frac{m\lambda_n }{\pi } \left(\int_{-\infty }^{\infty }\frac{
e^{ipx}F(n,p)}{2mE-p^2}\right)dp
\end{equation}
For $n=0$ we get the solution of the delta function that was discussed in details in \cite{Our,Our1}.
From eq.(\ref{GeneralPsiNon}), the bound state wave function in coordinate space is
\begin{equation}\label{GeneralPsi2}
\Psi_B(x)=\lambda_n \sum_{j=0}^{n}C^{n}_{j}
\Psi_B ^{(n-j)}(0)(-1)^{j}I_j(x,\Delta E_B),
\end{equation}
where $\Delta E_B<0$ is the binding energy, and
\begin{equation}\label{IkNon}
I_{j}(x,\Delta E_B)=\frac{m}{\pi }\left( P.V.\int_{-\infty }^{\infty }\frac{(ip)^{j}e^{ipx}%
}{2m\Delta E_B-p^2}dp\right) =\frac{\partial ^{j}I_{0}(x,\Delta E_B)}{\partial x^{j}%.
}.
\end{equation}
The above equations means that the expression of the wave function can be calculated from just calculating $I_{0}(x,\Delta E_B)$. This can be done using contour integral (see Figure 2 top panel). The result is
\begin{equation}\label{NonReIxspace}
 I_{0}(x,\Delta E_B) =-\sqrt{\frac{m}{-2 \Delta E_B}}\exp(-\sqrt{-2\Delta E_B m}|x|).
\end{equation}
\begin{figure}[tbh]
\begin{center}
\includegraphics[bb=280 28 400 400,scale=0.45]{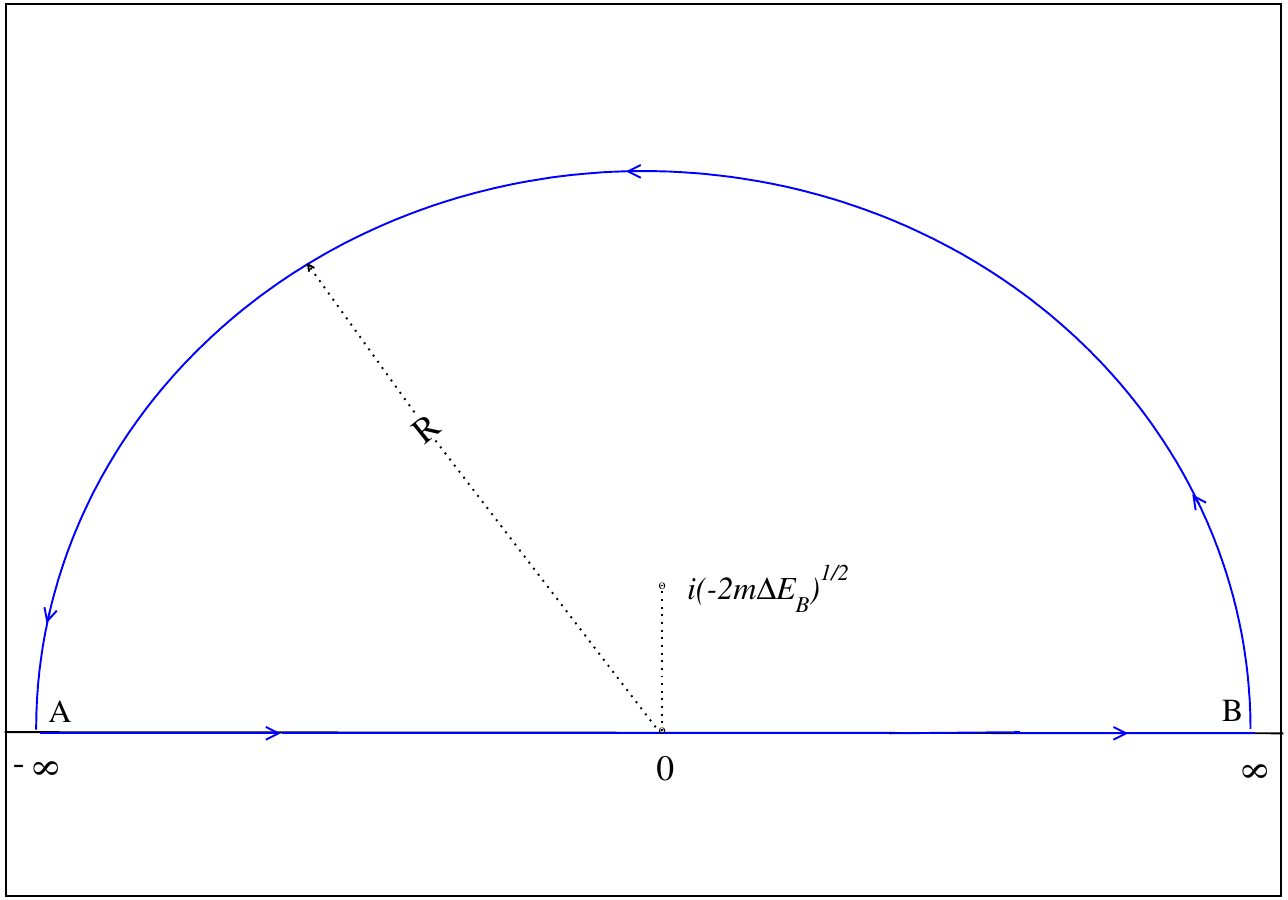}
\end{center}
\vspace{5mm}
\begin{center}
\includegraphics[bb=280 28 400 400,scale=0.45]{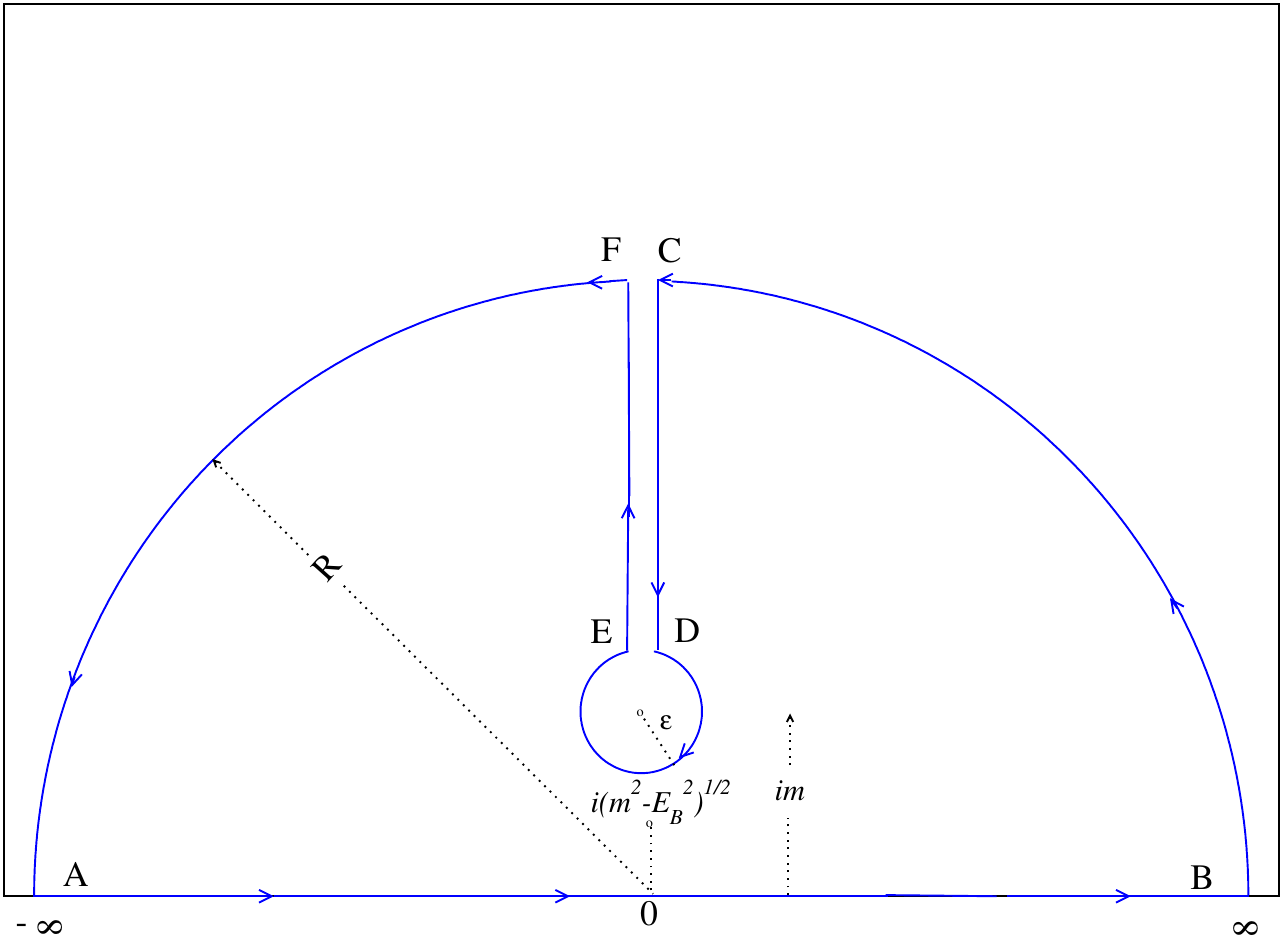}
\vspace{5mm}
\caption{\it The integration contours for obtaining the wave function of
the bound state. In the non-relativistic case, there is a pole inside the contour at $i\sqrt{-2m\Delta E_B}$, but no branch cut(top panel).
For relativistic case, there is a branch cut along the positive imaginary axis, starting at $p = im$, and there is a pole at $p = i \sqrt{m^2 - E_B^2}$ (bottom panel).}
\end{center}
\label{fig2}
\end{figure}
As it was explained in \cite{Widmer,Our1,Our2}, the non-relativistic problem needs to be regularized, therefore the expression of the wave function in eq.(\ref{GeneralPsiNon}) is considered as the unregularized expression of the bound state wave function. The regularization can be done by regularizing the integrals $I_{j}(x,\Delta E_B)$. For cutoff
regularization, the interval of the integral in eq.(\ref{IkNon}) is changed to $[-\Lambda,\Lambda]$, accordingly eq.(\ref{IkNon}) is written as
\begin{equation}\label{IkNonReg}
I_{j}(x,\Delta E_B,\Lambda)=\frac{m}{\pi }\left( P.V.\int_{-\Lambda }^{\Lambda }\frac{(ip)^{j}e^{ipx}%
}{2m\Delta E_B-p^2}dp\right) =\frac{\partial ^{j}I_{0}(x,\Delta E_B,\Lambda)}{\partial x^{j}}.
\end{equation}
The regularization includes the bare coupling constant as well. We define $I_j(\Delta E_B,\Lambda)\equiv I_j(0,\Delta E_B,\Lambda)$. From eq.(\ref{IkNonReg}), and for arbitrary $j$, the general expression of $I_j(\Delta E_B,\Lambda)$ can be obtained in terms of $\Lambda$. The result is
\begin{eqnarray}\label{IkHyper}
  I_{2j}(\Delta E_B,\Lambda)=\frac{(-1)^j}{2 \Delta E_B \pi (1+2j)}\Lambda^{1+2j}  _{2}F_1(1,j+\frac{1}{2};j+\frac{3}{2};\frac{\Lambda^2}{2\Delta E_B m}),
\end{eqnarray}
where $j=0,1,2,...$, and $_{2}F_1(a_1,a_2; a_3; z)$ is a hypergeometric function with one variable. On the other hand, it straightforward to prove that
\begin{equation}
     I_{2j+1}(\Delta E_B,\Lambda)=0,\hspace{20mm} j=0,1,2,..
\end{equation}
For $\Lambda\rightarrow\infty$, the asymptotic behavior of $I_{2j}(\Delta E_B,\Lambda)$ is given by the following relation
\begin{equation}\label{INonBinAsym}
    I_{2j}(\Delta E_B,\Lambda)\sim (-1)^{j-1}\frac{2m\Lambda^{2j-1}}{\pi(2j-1)}, \hspace{20mm} j=1,2,...
\end{equation}
At this point, the regularized form of the wave function can be introduced. It can be written as
\begin{equation}\label{RegBoundWF}
  \Psi (x)=\lim_{\Lambda\rightarrow \infty}\lambda_n (\Lambda)\sum_{j=0}^{n}C^{n}_{j}
\Psi ^{(n-j)}(0,\Lambda)(-1)^{j}I_j(x,\Delta E_B,\Lambda),
\end{equation}
The gap equation can be derived from the above equation. Substituting  $x=0$ in eq.(\ref{RegBoundWF}) leads to one algebraic equation, in addition, deriving eq.(\ref{RegBoundWF}) for one time, two times, until $n$-times at $x=0$ give additional $n$-equations. The $n+1$-equations are
\begin{eqnarray}\label{PsizeroSys}
  \Psi ^{(s)}(0,\Lambda)&=&\lambda_n (-1)^{n}\sum_{j=0}^{n}\Bigg(
C^{n}_{j} \frac{(-1)^{j+s}+(-1)^{j}+(-1)^{s}+1}{4}\Psi ^{(n-j)}(0,\Lambda)I_{j+s}\nonumber \\ &+&
 C^{n}_{j+1}
\frac{(-1)^{j+s}-(-1)^{j}+(-1)^{s}-1}{4}\Psi ^{(n-j-1)}(0,\Lambda)I_{j+s+1}\Bigg),
\end{eqnarray}
where $s=0,1,...n$, and $I_{m}$ is a short hand for $I_{m}(\Delta E_B,\Lambda)$. It is very important to distinguish between two cases; the first one, when the derivative of the $\delta$-function potential is an even number. In this case the wave function is either an even or an odd function. For an even wave function solution
\begin{equation}\label{EvenCon}
    \Psi ^{(1)}(0,\Lambda)=\Psi ^{(3)}(0,\Lambda)=...\Psi ^{(2n-1)}(0,\Lambda)=0,
\end{equation}
therefore using eq.(\ref{PsizeroSys}), the gap equation can be obtained from the roots of the following equation
\begin{equation}\label{SeqEqEven}
\left\vert
\begin{array}{cccccc}
-1+\lambda_n C^{n}_{n} I_{n}& \lambda_n  C^{n}_{n-2} I_{n-2}& ...& C^{n}_{n-2r} I_{n-2r} &...& I_{0}\\
\lambda_n C^{n}_{n}I_{n+2}& -1+C^{n}_{n-2}\lambda_n I_{n} &...& C^{n}_{n-2r} I_{n-2r+2} &...&  I_{2}\\
\vdots  & \vdots & \vdots &\vdots & \ddots &\vdots \\
\lambda_n C^{n}_{n}I_{n+2t}& \lambda_n C^{n}_{n-2}I_{n-2+2t} &...& C^{n}_{n-2r} I_{n-2r+2t} &...& -1+\lambda_n I_{2t}
\end{array}%
\right\vert =0,
\end{equation}
where $r,t=0,1,2,..n/2$. The values of $\Psi ^{(2)}(0,\Lambda),\Psi ^{(4)}(0,\Lambda),...\Psi ^{(2n)}(0,\Lambda)$ can be calculated in terms of $\Psi ^{(0)}(0,\Lambda)$ by solving eqs.(\ref{PsizeroSys}), and using one of the $n$ roots from eq.(\ref{SeqEqEven}) at a time.
For the odd wave function solution,
\begin{equation}\label{OddCon}
    \Psi ^{(0)}(0,\Lambda)=\Psi ^{(2)}(0,\Lambda)=...\Psi ^{(2n)}(0,\Lambda)=0,
\end{equation}
therefore using eq.(\ref{PsizeroSys}), the gap equation can be obtained from the roots of the following equation
\begin{equation}\label{SeqEqOdd}
\left\vert
\begin{array}{cccccc}
1+\lambda_n C^{n}_{n-1} I_{n}& \lambda_n  C^{n}_{n-3} I_{n-2}& ...& C^{n}_{n-2r-1} I_{n-2r} &...& C^{n}_{1} I_{2}\\
\lambda_n C^{n}_{n-1}I_{n}& 1+C^{n}_{n-3}\lambda_n I_{n+2} &...& C^{n}_{n-2r-1} I_{n-2r+2} &...& C^{n}_{1} I_{4}\\
\vdots  & \vdots & \vdots &\vdots & \ddots &\vdots  \\
\lambda_n C^{n}_{n-1}I_{n+2t}& \lambda_n C^{n}_{n-3}I_{n-2+2t} &...& C^{n}_{n-2r-1} I_{n-2r+2t} &...& 1+\lambda_n C^{n}_{1} I_{2t}
\end{array}%
\right\vert =0,
\end{equation}
where $r,t=0,1,..(n-2)/2$. The values of $\Psi ^{(3)}(0,\Lambda), \Psi ^{(5)}(0,\Lambda),...\Psi ^{(2n-1)}(0,\Lambda)$ can be calculated in terms of $\Psi ^{(1)}(0,\Lambda)$ by solving eqs.(\ref{PsizeroSys}), and using one of the roots $n-1$ from eq.(\ref{SeqEqOdd}), one root at a time. In total, we have $2n-1$ solutions for the even derivative case.

The second case is when the derivative of the $\delta$-function potential is an odd number. In this case the solution is neither a n odd nor an even function
\begin{eqnarray}\label{SeqNOE}
\left\vert
\begin{array}{cccccc}
 -1/\lambda_n& A^n_{00}&\cdots & A^{n}_{0r}&\cdots& A^{n}_{0n}\\
 A^{n}_{10}& -1/\lambda_n& \cdots & A^{n}_{1r}&\cdots& A^{n}_{1n}\\
\vdots  & \vdots & \ddots &\vdots & \vdots &\vdots \\
A^{n}_{s0}&\cdots&\cdots&A^{n}_{sr}&\cdots&A^{n}_{sn}\\
\vdots  & \vdots & \vdots &\vdots & \ddots &\vdots \\
A^{n}_{n0}&\cdots&\cdots &A^{n}_{nr}&\cdots&-1/\lambda_n
\end{array}
\right\vert =0%,
\end{eqnarray}
where
\begin{eqnarray}\label{AnEq}
  A^{n}_{sr} =\frac{(-1)^n}{2}C_{n-r}^n I_{n-r+s}(\Delta E_B,\Lambda) \left((-1)^{n-r}+(-1)^s\right).
\end{eqnarray}
From eq.(\ref{SeqEqOdd}), it is clear that there are $n+1$ gap equations for the $n$-odd case.

To be a physical state, the wave function for the bound state must be normalizable. From eq.(\ref{FnpEq}) and eq.(\ref{GeneralPsiNon}) we get
\begin{eqnarray}\label{NormaNonGen}
  \int_{-\infty}^{\infty}|\Psi_B(x)|^2 dx&=&\lim_{\Lambda\rightarrow \infty}\frac{2m^2\lambda_n(\Lambda)^2}{\pi}\sum_{r=0}^{2n} \sum_{j=0}^{n}i^r C^{n}_{j}C^{n}_{r-j}(-1)^{j}\nonumber\\ &\times&\Psi_B ^{(n-j)}(0,\Lambda)\Psi_B ^{(n-r+j)} (0,\Lambda) \int_{-\Lambda}^{\Lambda}\frac{p^{2r}}{(2m \Delta E_B-p^2)^2} dp
 \nonumber\\ & =&1.
\end{eqnarray}
At first glance, it seems that the wave function is not normalizable because
\begin{equation}\label{div}
   \lim_{\Lambda\rightarrow\infty} \int_{-\Lambda}^{\Lambda} \frac{p^{r}}{(2m \Delta E_B-p^2)^2} dp\rightarrow\infty,\hspace{20mm} r= 1,2,...
\end{equation}
However further analysis shows that this is not the case, as quantities like $\lambda_n \Psi_B ^{(m)}$  goes to zero fast enough as $\Lambda\rightarrow \infty$ for all $m<n$ such that the integrals in eq.(\ref{NormaNonGen}) converges. This can be very well demonstrated in section 3.

The scattering wave function for the non-relativistic case can be studied using the following ansatz
\begin{equation}\label{GeneralAnsatzNon}
  \tilde{\Psi}_E(p) =A\delta(p-\sqrt{2m\Delta E})+B\delta(p+\sqrt{2m\Delta E})+\tilde{\Phi}_E(p), \hspace{20 mm} \Delta E=\frac{k^2}{2m},
\end{equation}
where $A$ and $B$ are arbitrary constants that will be defined later.
To calculate the scattering states, the expression of  $\Phi _{E}(x)$ must be calculated first. Substituting for $\tilde{\Psi}_E(p)$  from eq.(\ref{GeneralAnsatzNon}) into eq.(\ref{pSpaceNon1}), and then solving for   $\tilde{\Phi}_E(p)$ we get
\begin{equation}\label{PhiNonP}
\tilde{\Phi}_E(p)=\frac{2m \lambda_n
}{2m \Delta E-p^2}\sum_{j=0}^{n} (-1)^{n}(-ip)^{j}C^{n}_{j}\left( \frac{A+(-1)^{n-j}B}{2\pi }(i k)^{n-j}+\Phi^{n-j} _{E}(0)\right).
\end{equation}
In $x$-space,
\begin{equation}\label{GeneralNonXScat}
   \Phi_E(x)=\lambda_n\sum_{j=0}^{n} (-1)^{j+n}C^{n}_{j}I_{2j+1}(x,\Delta E)\left( \frac{A+(-1)^{n-j}B}{2\pi }(i k)^{n-j}+\Phi^{n-j} _{E}(0)\right),
\end{equation}
where
\begin{equation}\label{IkNonScat}
I_{j}(x,\Delta E)=\frac{m}{\pi }\left( P.V.\int_{-\infty }^{\infty }\frac{(ip)^{j}e^{ipx}%
}{2m\Delta E-p^2}dp\right) =\frac{\partial ^{j}I_{0}(x,\Delta E)}{\partial x^{j}%.
}.
\end{equation}
By using cutoff momentum in the expression of $I_{j}(x,\Delta E)$ in eq.(\ref{IkNonScat}), the function $\Phi_E(x)$ is regularize to $\Phi_E(x,\Lambda)$. This leads to the following $n+1$ equations
\begin{eqnarray}\label{PhiZeroEqs}
\Phi_E ^{(s)}(0,\Lambda)&=&\lambda_n (-1)^{n}\sum_{j=0}^{n}\Bigg(
C^{n}_{j} \frac{(-1)^{j+s}+(-1)^{j}+(-1)^{s}+1}{4}I_{j+s}(\Delta E,\Lambda)\nonumber\\&\times& \left( \frac{A+(-1)^{n-j}B}{2\pi }(i k)^{n-j}+\Phi^{n-j} _{E}(0,\Lambda)\right)\nonumber \\ &+&
 C^{n}_{j+1}
\frac{(-1)^{j+s}-(-1)^{j}+(-1)^{s}-1}{4} I_{j+s+1}(\Delta E,\Lambda)\nonumber\\&\times&\left( \frac{A+(-1)^{n-j-1}B}{2\pi }(i k)^{n-j-1}+\Phi^{n-j-1} _{E}(0,\Lambda)\right)\Bigg),
\end{eqnarray}
where $s=0,1,...n$, and $I_j(\Delta E,\Lambda)\equiv I_j(0,\Delta E\Lambda,)$. From eq.(\ref{IkNonScat}), it can be proved that $ I_0(0,\Delta E)=0$. On the other hand, introducing the cutoff leads to the following generalized relation
\begin{equation}
    I_{2j}(\Delta E,\Lambda)=\frac{2m(-1)^{j}}{\pi}\left(k^{2j-1}\text{arccoth}\left(\frac{\Lambda}{k} \right)-k^{2j-2}\sum_{r=0}^{j-1}\frac{\Lambda^{2r+1}}{(2r+1)k^{2r}}\right),
\end{equation}
again here
\begin{equation}
     I_{2j+1}(\Delta E,\Lambda)=0,\ \ \ \ j=0,1,2,...
\end{equation}
For $j=0$, we get
\begin{equation}
    I_0(\Delta E,\Lambda)= \frac{2m}{k\pi}\text{arccoth}\left(\frac{\Lambda}{k} \right),
\end{equation}
which leads to
\begin{equation}\label{Izero}
   \lim_{ \Lambda\rightarrow\infty}I_0(\Delta E,\Lambda)=0
\end{equation}
For $\Lambda\rightarrow\infty$, the asymptotic of behavior $I_{2j}(\Delta E,\Lambda)$ is given by the following relation
\begin{equation}\label{INon}
    I_{2j}(\Delta E,\Lambda)\sim (-1)^{j-1}\frac{2m\Lambda^{2j-1}}{\pi(2j-1)}, \hspace{20 mm} j=1,2,...
\end{equation}

At this stage, there is everything needed to calculate the scattering wave function by using eqs.(\ref{PhiZeroEqs}).
\section{The solution for the non-relativistic $\delta^{(2)}$-potential}
For this case, the potential is an even function,  therefore the solution is either an even or an odd function. For the even function solution,  eq.(\ref{SeqEqEven}) is applicable, accordingly for $n=2$, the gap equations are
\begin{equation}\label{Gap2Der}
  \frac{1}{\lambda_2(\Lambda)} =I_{2}(\Delta E_B,\Lambda)\pm \sqrt{I_{0}(\Delta E_B,\Lambda)I_{4}(\Delta E_B,\Lambda)}=\frac{1}{\lambda_2^{(1,2)}(\Lambda)} .
\end{equation}
The regularized form of the wave function is
\begin{equation}\label{RegBoundWF}
  \Psi (x)=\lim_{\Lambda\rightarrow \infty}\frac{m}{\pi}\lambda_2 (\Lambda)\int_{-\Lambda}^{\Lambda}\frac{\Psi^{(2)}_B(0,\Lambda)-p^{2}\Psi_B(0,\Lambda)}{2m \Delta E_B-p^2}e^{ipx} dp.
\end{equation}
By solving eqs.(\ref{PsizeroSys}), the value of $\Psi_B^{(2)}(0,\Lambda)$ can be obtained in terms of $\Psi_B(0,\Lambda)$. The result is
\begin{equation}
    \Psi_B^{(2)}(0,\Lambda)=\pm\Psi_B(0,\Lambda)\sqrt{\frac{I_{4}(\Delta E_B,\Lambda)}{I_{0}(\Delta E_B,\Lambda)}},\hspace{20mm}\lambda_2 (\Lambda)=\lambda_2^{(1,2)}(\Lambda)
\end{equation}
from the asymptotic behavior of $I_{2j}(\Delta E_B,\Lambda)$ when $\Lambda\rightarrow\infty$,  we know that $I_{4}(\Delta E_B,\Lambda)\sim \Lambda^{3}$, while $I_{0}(\Delta E_B,\Lambda)$ is a constant.

The wave function must be normalizable. From eq.(\ref{NormaNonGen}), the normalization condition is
\begin{equation}\label{NormSecond1}
   \lim_{\Lambda\rightarrow \infty} \frac{2m^{2}\lambda_2(\Lambda)^2}{\pi}\int_{-\Lambda}^{\Lambda}\frac{p^{4}\Psi_B(0,\Lambda)^2-2p^{2}\Psi_B^{(2)}(0,\Lambda)\Psi_B(0,\Lambda)+\Psi_B^{(2)}(0,\Lambda)^2}{(2m \Delta E_B-p^2)^2} dp=1.
\end{equation}
In the above expression,
\begin{eqnarray}\label{DiveQauntities1}
    \int_{-\Lambda}^{\Lambda}dp\frac{p^{4}}{(2m\Delta E_B-p^2)^2} &\sim& \Lambda, \ \ \ \ \ \lim_{\Lambda\rightarrow \infty}\int_{-\Lambda}^{\Lambda}dp\frac{p^{2}}{(2m\Delta E_B-p^2)^2}=\sqrt{\frac{\pi^2}{-8m\Delta E_B}}, \nonumber\\ \lim_{\Lambda\rightarrow \infty}\int_{-\Lambda}^{\Lambda}dp\frac{1}{(2m\Delta E_B-p^2)^2}&=&\frac{\pi}{2(-2m\Delta E_B)^{3/2}},
\end{eqnarray}
on the other hand, for this case
\begin{equation}\label{DiveQauntities2}
    \lambda_2(\Lambda)^2 \sim   \Lambda^{-3}, \hspace{20mm}\Psi_B^{(2)}(0,\Lambda)\sim  \Lambda^{3/2} \Psi_B(0,\Lambda).
\end{equation}
The above relations mean that the leading term in eq.(\ref{NormSecond1}) is the one with $\Psi_B^{(2)}(0,\Lambda)^2$. All the other terms vanish as the cutoff is removed. Therefore we reach to the important result that $\lambda_2(\Lambda) \Psi_B^{(2)}(0,\Lambda)$ is a finite quantity, although $\Psi_B^{(2)}(0,\Lambda)\rightarrow \infty$ as $\Lambda\rightarrow\infty$. The normalization condition gives
\begin{equation}\label{NonRelNor}
 \lim_{\Lambda\rightarrow \infty} \lambda_2(\Lambda)\Psi_B^{(2)}(0,\Lambda)=\pm\frac{ (-2\Delta E_B )^{3/4}}{m^{1/4}}.
\end{equation}
Finally, the bound state for this case can be written as
\begin{eqnarray}\label{NonRelWFfinal1}
    \Psi_B (x)&=&\pm\frac{\varkappa^{3/2}}{m}\Bigg(-\frac{m}{\varkappa}\exp(-\varkappa|x|)\nonumber\\
    &\pm& \lim_{\Lambda\rightarrow \infty}\frac{m}{\pi}  \sqrt{\frac{I_0(\Delta E_B,\Lambda)}{I_4(\Delta E_B,\Lambda)}}\int_{-\Lambda }^{\Lambda }\frac{p^2 e^{ipx}}{2m\Delta E_B-p^2}dp\Bigg),
\end{eqnarray}
The integral in the second term in the above equation has an extremum at $x=0$ proportional to $\Lambda$ as $\Lambda\rightarrow \infty$. Nevertheless, the extremum value times $\sqrt{I_0/I_4}$ is suppressed because $I_4(E_B,\Lambda)\sim \Lambda^3$ as $\Lambda\rightarrow \infty$. Therefore the second term can be ignored relative to the first term. However, we can not simply say that the second term is zero, because eqs.(\ref{PsizeroSys}) have to be satisfied.
\begin{figure}[tbh]
\begin{center}
\includegraphics[bb=280 28 400 400,scale=0.46]{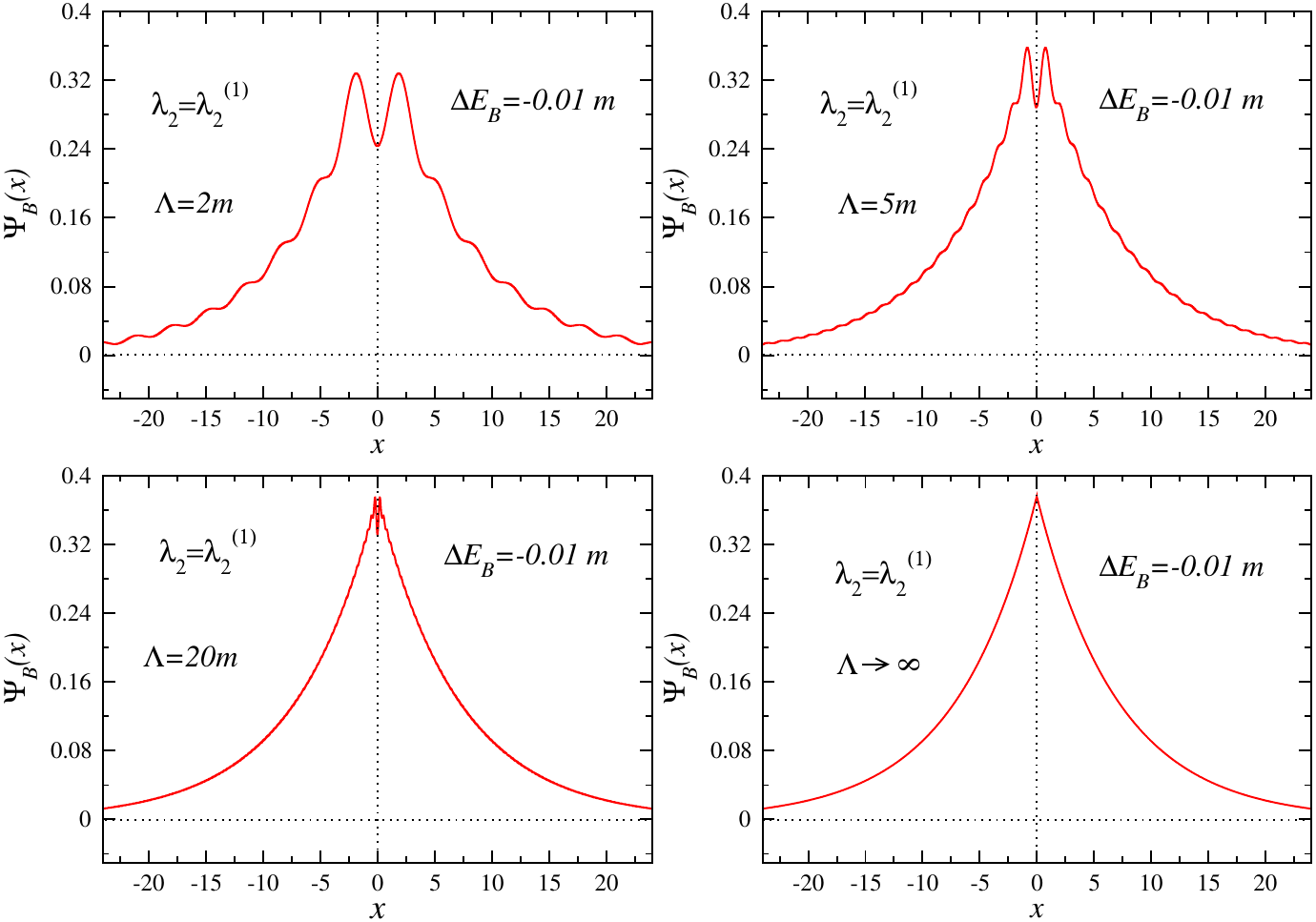}
\end{center}
\vspace{5mm}
\begin{center}
\includegraphics[bb=280 28 400 400,scale=0.46]{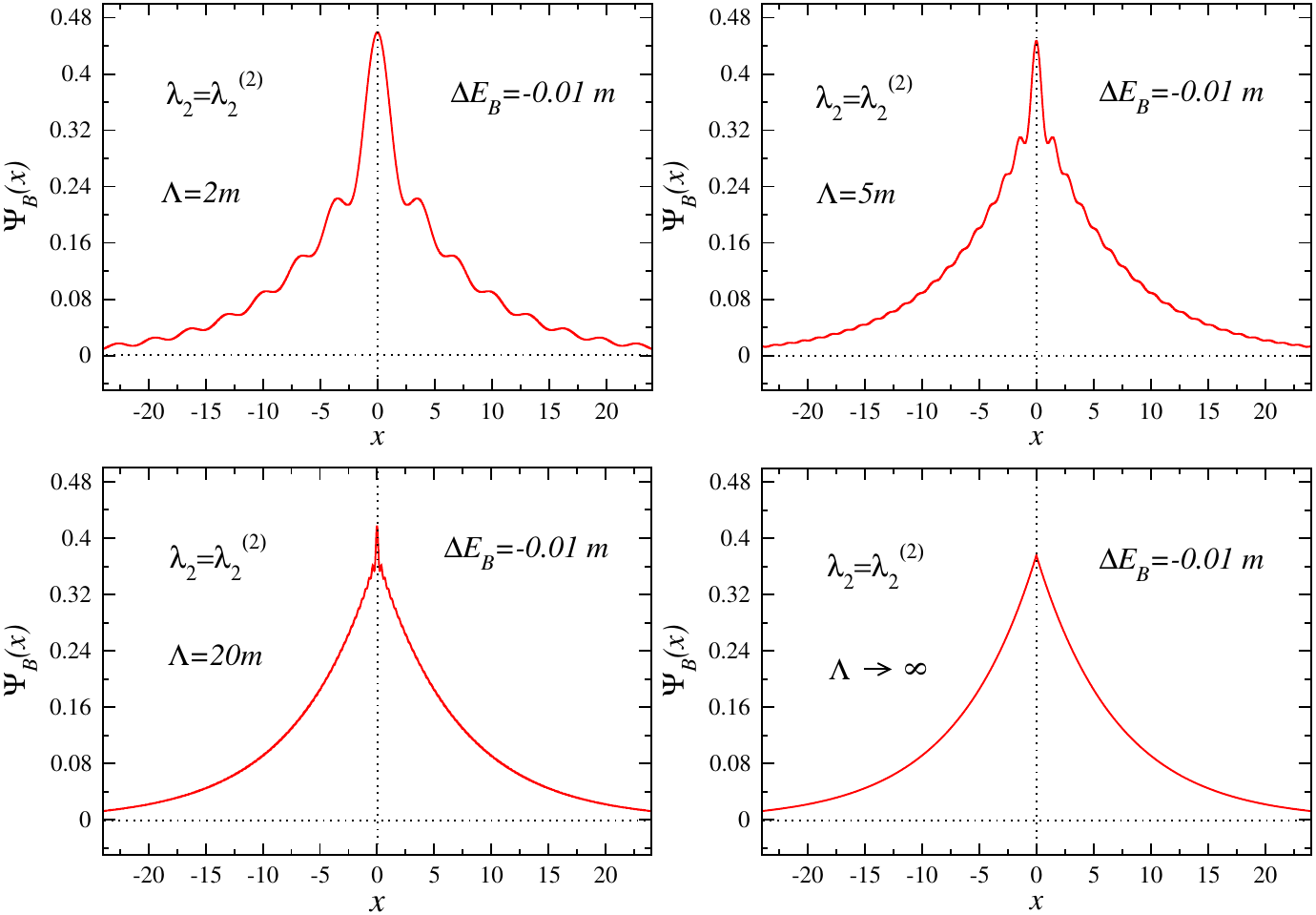}
\vspace{5mm}
\caption{\it Bound state wave function in coordinate space for the non-relativistic $\delta^{(2)}$-function potential case. The value of the binding energy is $\Delta E_B =-0.01 m$, with different values of $\Lambda=2m, 5m,20m$, and $\infty$. For $\lambda_2=\lambda_2^{(1)}$ (top four figures), the curves are different than the analogs ones for the case $\lambda_2=\lambda_2^{(2)}$ (bottom four figures), however, the difference diminish with the increase of $\Lambda$. The wave function in all the figures is an even function for any value of $\Lambda$.}
\end{center}
\label{fig3}
\end{figure}
The essence of calculating the scattering states is to calculate the value of $\Phi_E ^{(s)}(0)$. This can be done by using eqs.(\ref{PhiZeroEqs}) for the case $n=2$, which lead to the values of $\Phi_E (0)$, $\Phi_E ^{(1)}(0)$ and $\Phi_E ^{(2)}(0)$ in terms of $I_{2j}(\Delta E_B,\Lambda)$, $I_{2j}(\Delta E,\Lambda)$ ($j=0,1,2$), and $ \lambda_2(\Lambda)$. For the even solution case, and as $\Lambda\rightarrow\infty$, the calculations lead to
\begin{eqnarray}\label{Phi0Non2}
   \lim_{\Lambda\rightarrow\infty}\lambda_2 (\Lambda)\left(-\frac{A+B}{2\pi}k^2+\Phi^{(2)} _{E}(0,\Lambda)\right)&=&\frac{A+B}{2\pi(I_{0}(\Delta E_B)-I_{0}(\Delta E))},  \nonumber\\ -2\lambda_2(\Lambda) \left(\frac{A-B}{2\pi}ik+\Phi^{(1)} _{E}(0,\Lambda)\right)&\sim&  \frac{1}{\Lambda^{3/2}}\nonumber\\\lambda_2(\Lambda) \left(\frac{A+B}{2\pi}+\Phi_{E}(0,\Lambda)\right)&=&\frac{A+B}{2\pi\sqrt{I_{0}(\Delta E_B,\Lambda)I_{4}(\Delta E_B,\Lambda)}}\nonumber\\&\sim&  \frac{1}{\Lambda^{3/2}}.
\end{eqnarray}
By using eqs.(\ref{Phi0Non2}) into eq.(\ref{GeneralNonXScat}), $\Phi_E (x)$  is
\begin{eqnarray}\label{PhideltaDerxFinal}
 \Phi_{E}(x)&=& (A+B)\Bigg(\frac{m}{2\pi^2I_0(\Delta E_B)}  \int_{-\infty }^{\infty }\frac{e^{ipx}dp}{2m \Delta E-p^2}\nonumber\\
 &-&\lim_{\Lambda\rightarrow\infty}\frac{2m}{\pi}\lambda_2(\Lambda) \left(\frac{A-B}{2\pi}ik+\Phi^{(1)}_{E}(0,\Lambda)\right)\int_{-\Lambda }^{\Lambda}\frac{ip \ e^{ipx}dp}{2m \Delta E-p^2}\Bigg),
 \nonumber\\
 &+&\lim_{\Lambda\rightarrow\infty}\frac{m}{\pi}\lambda_2(\Lambda) \left(\frac{A+B}{2\pi}+\Phi_{E}(0,\Lambda)\right)\int_{-\Lambda }^{\Lambda}\frac{-p^2 \ e^{ipx}dp}{2m \Delta E-p^2}\Bigg), \nonumber\\ \lambda_2(\Lambda)&=& \lambda_2^{(1,2)}(\Lambda).
\end{eqnarray}
In the expression of $\Phi _{E}(x)$, the second and third terms vanish as $\Lambda\rightarrow\infty$. However, at the contact point when $x=0$ these terms cannot be ignored, where  eqs(\ref{PhiZeroEqs}) are  satisfied. The first integral can be calculated using contour integral (see Figure 4 top panel).
\begin{figure}[tbh]
\begin{center}
\includegraphics[bb=280 28 400 400,scale=0.45]{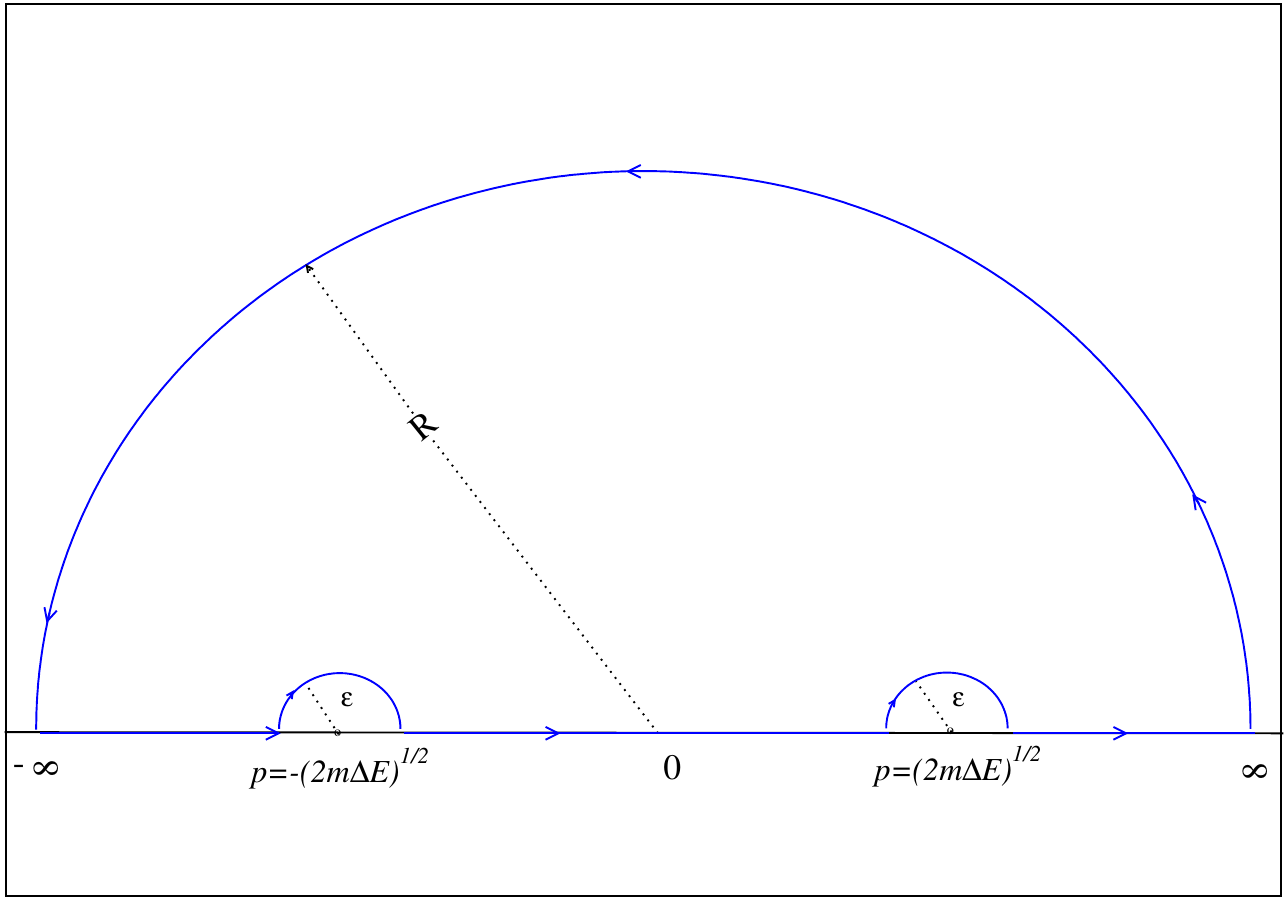}
\end{center}
\vspace{8mm}
\begin{center}
\includegraphics[bb=280 28 400 400,scale=0.45]{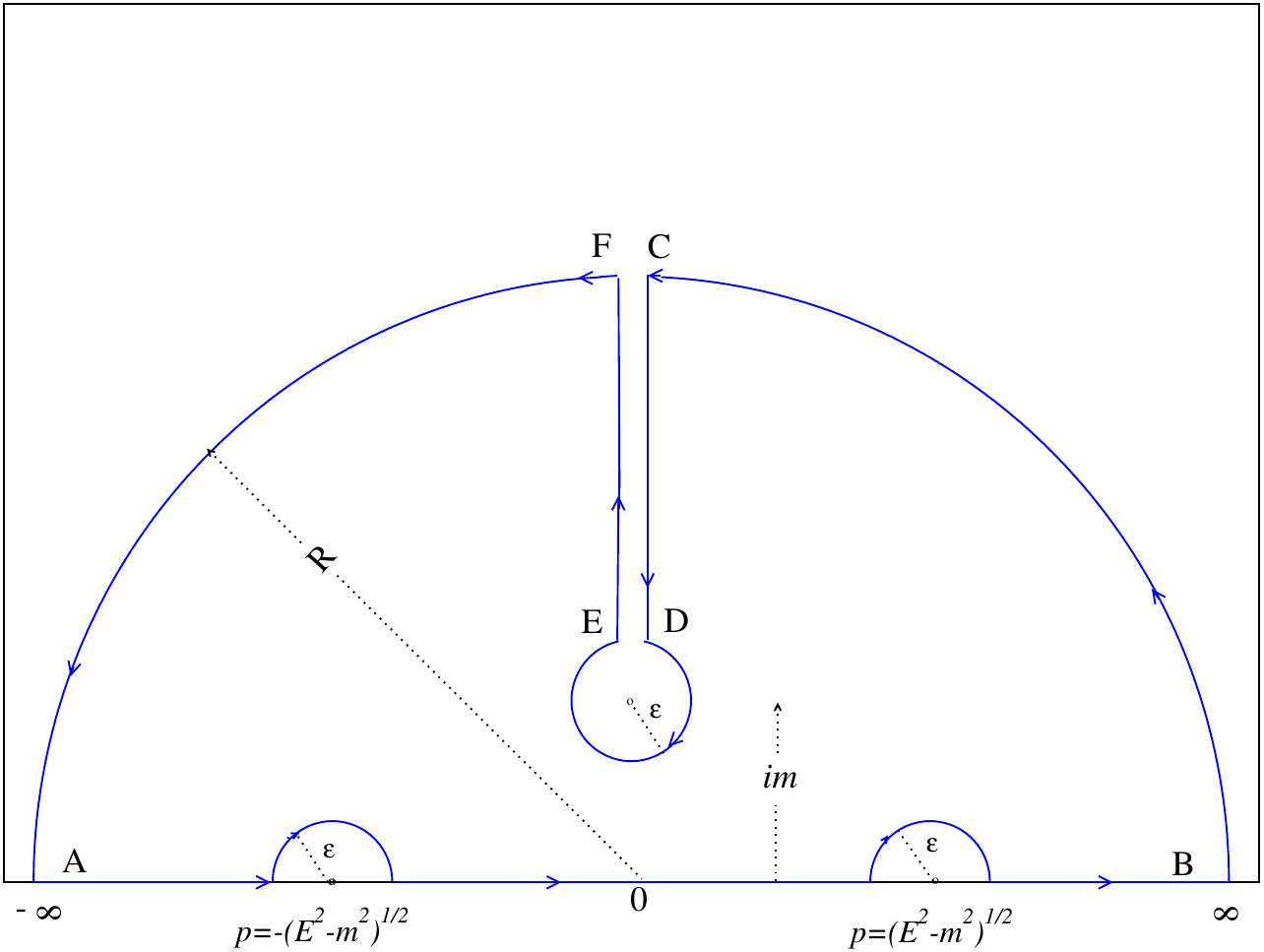}
\vspace{5mm}
\caption{\it The integration contours for obtaining the wave function of
scattering states. In the non-relativistic case, there are two  poles on the reals axis  at $\pm\sqrt{2m\Delta E}$, but no branch cut(top panel).
For relativistic case, there is a branch cut along the positive imaginary axis, starting at $p = im$, and there are two  poles on the real axis at $p = \pm\sqrt{E^2 - m^2}$ (bottom panel).}
\end{center}
\label{fig4}
\end{figure}
By using eq.(\ref{Izero}),  eq.(\ref{PhideltaDerxFinal}), into eq.(\ref{GeneralAnsatzNon}), the scattering wave function in $x$-space is
\begin{eqnarray}\label{PsiXdeltaFirstNon}
\Psi_E(x)&=& A e^{ikx} +B e^{-ikx}+
\frac{(A+B)}{ I_0(\Delta E_B)} \frac{m \sin(k |x|)}{k} \nonumber \\ &-&4\pi\lim_{\Lambda\rightarrow\infty} \lambda_2(\Lambda) \left(\frac{A-B}{2\pi}ik +\Phi^{(1)} _{E}(0,\Lambda)\right)I_1(x,\Delta E,\Lambda)\nonumber\\&+&2\lambda_2(\Lambda)\pi \left(\frac{A+B}{2\pi}+\Phi_{E}(0,\Lambda)\right)I_2(x,\Delta E,\Lambda), \nonumber\\ \lambda_2(\Lambda)&=& \lambda_2^{(1,2)}(\Lambda).
\end{eqnarray}
With the exception of the second and third terms, this is exactly like the scattering wave function for the $\delta$-potential \cite{Our,Our1,Our2}. Again here the  renormalized coupling constant can be defined as
\begin{equation}\label{LambdaEB}
    \lambda(\Delta E_B)\equiv \lambda= \frac{1}{I_0(\Delta E_B)}= -\sqrt{\frac{-2\Delta E_B}{m}}\Rightarrow  \ \ \ \Delta E_B=-\frac{m \lambda^2 }{2},
\end{equation}
It is important to mention here that for both cases of $\lambda_2^{(1,2)}(\Lambda)$, there is only attractive scattering states given by eq.(\ref{PsiXdeltaFirstNon}) with $ \lambda(\Delta E_B)<0$. This means that the regularization does not lead to a repulsive $\delta^{(2)}$-function potential.

The reflected wave function in the region $I$ to the left of the contact point, and transmitted wave function in the region $II$ to the right of the contact point are defined as
\begin{equation}\label{ReflDelNon}
\Psi_I(x) = \exp(i k x) + R(k) \exp(- i k x),\ \ \ \Psi_{II}(x) = T(k) \exp(i k x).
\end{equation}
From the above two equations and from eq.(\ref{PsiXdeltaFirstNon}) we get
\begin{eqnarray}
R(k) = - \frac{im\lambda}
{k + i m \lambda}, \quad
T(k) = \frac{k}{k + i m \lambda},\nonumber\\
A= \frac{2k + i m \lambda}{2k + 2i m \lambda}, \ \ \ \ \quad B=- \frac{i m \lambda}{2k + 2i m \lambda}\Rightarrow A+B=T(k),
\end{eqnarray}
where $R(k)$ is the reflection coefficient, and $T(k)$ is the transition coefficient.

To verify that the resulting system is self-adjoint, we have to prove that the scalar product of the bound state with a scattering state vanishes, or
\begin{equation}\label{ScNonEEB}
    \langle \Psi_B|\Psi_E\rangle=0,\ \ \ \ \ \ \ \lambda_2(\Lambda)= \lambda_2^{(1,2)}(\Lambda),
\end{equation}
we must also prove that the scalar product of a scattering state with energy $E'$ with another scattering state with energy $E$ gives
\begin{equation}\label{ScNonEpE}
\langle \Psi_{E'}|\Psi_E\rangle \sim
\delta(\sqrt{2m \Delta E} - \sqrt{ 2m \Delta E'}),\ \ \ \ \ \ \ \ \  \  \lambda_2(\Lambda)= \lambda_2^{(1,2)}(\Lambda).
\end{equation}
The calculation for proving this are lengthy, however the approached used here is similar to the one that is discussed in details in \cite{Our2} appendix B.
\subsection{The odd wave function solution}
For the odd function solution, $\Psi_B(0,\Lambda)=\Psi_B^{(2)}(0,\Lambda)=0$, for this case  eq.(\ref{SeqEqOdd}) is applicable, accordingly for $n=2$, the gap equation is
\begin{equation}\label{Gap2DerOdd}
  \frac{1}{\lambda_2(\Lambda)} =-2I_{2}(\Delta E_B,\Lambda).
\end{equation}
The regularized form of the wave function for this case is
\begin{equation}\label{RegBoundWF}
  \Psi_B(x)=\lim_{\Lambda\rightarrow \infty}\frac{m}{\pi}\lambda_2 (\Lambda)\int_{-\Lambda}^{\Lambda}\frac{-2ip\Psi_B^{(1)}(0,\Lambda)}{2m \Delta E_B-p^2} dp.
\end{equation}
The wave function must be normalizable. From eq.(\ref{NormaNonGen}), the normalization condition is
\begin{equation}\label{NormSecond1}
   \lim_{\Lambda\rightarrow \infty} \frac{2m^{2}\lambda_2(\Lambda)^2}{\pi}\int_{-\Lambda}^{\Lambda}\frac{4p^{2}\Psi_B^{(1)}(0,\Lambda)^2}{(2m \Delta E_B-p^2)^2} dp=1.
\end{equation}
In the above expression,
\begin{eqnarray}\label{DiveQauntities1}
     \lim_{\Lambda\rightarrow \infty}\int_{-\Lambda}^{\Lambda}dp\frac{p^{2}}{(2m\Delta E_B-p^2)^2}=\sqrt{\frac{\pi^2}{-8m\Delta E_B}},
\end{eqnarray}
on the other hand, for this case, and as $\Lambda\rightarrow \infty$,
\begin{equation}\label{DiveQauntities2}
    \lambda_2(\Lambda) \sim   \Lambda^{-1}.
\end{equation}
As a result the wave function is normalizable under the condition
\begin{equation}
   \lim_{\Lambda\rightarrow \infty}\lambda_2(\Lambda) \Psi_B^{(1)}(0,\Lambda)=\pm \left(\frac{-\Delta E_B}{8m^3}\right)^{1/4}.
\end{equation}
Accordingly, the wave function for the bound state for this case is
\begin{equation}\label{NonBoundOdd}
  \Psi_B(x)=(\varkappa)^{1/2}\text{sgn}(x) \exp(-\varkappa |x|)
\end{equation}
For the scattering states, the expressions of $\Phi_E (0)$, $\Phi_E ^{(1)}(0)$ and $\Phi_E ^{(2)}(0)$ can be calculated this time too using $\lambda(\Lambda)$ from eq.(\ref{Gap2DerOdd}). As $\Lambda\rightarrow\infty$, the calculations lead to
\begin{eqnarray}\label{Phi0Non2}
   \lambda_2 (\Lambda)\left(-\frac{A+B}{2\pi}k^2+\Phi^{(2)} _{E}(0,\Lambda)\right)\sim \Lambda \rightarrow \infty, \nonumber\\ -2\lambda_2(\Lambda) \left(\frac{A-B}{2\pi}ik+\Phi^{(1)} _{E}(0,\Lambda)\right)\sim  \Lambda^0, \nonumber\\
   \lambda_2(\Lambda) \left(\frac{A+B}{2\pi}+\Phi_{E}(0,\Lambda)\right)\sim  \frac{1}{\Lambda}.
\end{eqnarray}
The above equations means that $\Phi_E (x)$ is divergent for any value of $x$. Therefore the problem is nonrenormalizable for the odd bound state case.
\section{The Bound State of the Relativistic Problem }
The relativistic time -independent Schr\"odinger equation for the $\delta^{(n)}$- potential is
\begin{equation}\label{SchNDRel1}
\sqrt{p^{2}+m^{2}}\Psi (x)+\lambda_n \delta ^{(n) }(x)\Psi (x)=E\Psi(x)
\end{equation}
In $p$-space, the above equation can be written as
\begin{equation}\label{SchNDRel2}
\sqrt{p^{2}+m^{2}}\widetilde{\Psi }(p)+\lambda_n \int_{-\infty }^{\infty
}\delta ^{(n)}(x)\Psi (x)e^{-ipx}dx=E\widetilde{\Psi}(p).
\end{equation}
By using the results from the non-relativistic case, the wave function in $x$-space is
\begin{equation}\label{GeneralPsiRel}
\Psi(x)=\frac{\lambda_n }{2\pi } \int_{-\infty }^{\infty }\frac{
e^{ipx}F(n,p)}{E-\sqrt{p^{2}+m^{2}}}dp,
\end{equation}
where $F(n,p)$ is defined by eq.(\ref{FnpEq}). The cases $n=0,1$  were discussed in details \cite{Our,Our2}. For the bound state, eq.(\ref{GeneralPsiRel}) can be written as
\begin{equation}\label{GeneralPsi2}
\Psi (x)=\lambda_n \sum_{j=0}^{n}C_{j}^{n}
 \Psi_B ^{(n-k)}(0)(-1)^{j}I_j(x,E_B),
\end{equation}
where
\begin{equation}
I_j(x,E_B)=\frac{1}{2\pi }\left( P.V.\int_{-\infty }^{\infty }\frac{p^{j}e^{ipx}}{E_B-\sqrt{p^{2}+m^{2}}}dp\right).
\end{equation}
From the above equation, it is straightforward to prove that
\begin{equation}
I_{j}(x,E_B)=\frac{\partial ^{j}I_{0}(x,E_B)}{\partial x^{j}},
\end{equation}
this means that the problem is reduced to obtaining the expression of $I_{0}(x,E_B)$. Here, there are three possible cases that decide the expression of $I_{0}(x,E_B)$; a bound state when $0 < E_B < m$, strong bound state when $-m < E_B < 0$, and ultra-strong bound state when $ E_B < -m$. All these cases $I_{0}(x,E_B)$ can be obtained elegantly by using contour integral \cite{Our,Our2}(see Figure 1 bottom panel). For the bound state when $0 < E_B < m$, the integrand has a pole at $p = i \sqrt{m^2 - E_B^2}$, which is enclosed by $\Gamma$, as well as a branch cut along the positive imaginary axis starting at $p = i m$. Accordingly
\begin{equation}\label{boundstateI}
I_{0}(x,E_B)= - \frac{1}{\pi} \int_m^\infty d\mu
\frac{\sqrt{\mu^2 - m^2}}{E_B^2 - m^2 - \mu^2} \exp(- \mu |x|) -
\frac{E_B \exp(- \sqrt{m^2 - E_B^2} |x|)}{\sqrt{m^2 - E_B^2}}.
\end{equation}
For strong bound state when $-m<E_B<m$, and ultra bound state when $-m>E_B$, the pole inside the contour gives no residue. Therefor
\begin{equation}\label{StboundstateI}
I_{0}(x,E_B)= - \frac{1}{\pi} \int_m^\infty d\mu
\frac{\sqrt{\mu^2 - m^2}}{E_B^2 - m^2 - \mu^2} \exp(- \mu |x|),
\end{equation}
and $\Psi_B(x)$ takes the following neat general expression
\begin{equation}\label{}
\Psi_B(x)=\lambda_n\left( \int_{m}^{\infty }d\mu\frac{\sqrt{
u^{2}-m^{2}}}{E_B^2 - m^2 + \mu^2}F(n,iu)e^{-u|x|}\right).
\end{equation}
The expression for $\Psi_B(x)$ obtained by using eq.(\ref{boundstateI}) or eq.(\ref{StboundstateI}) leads to a wave function that is non-normalizable for $n\geq 1$. Therefore, the cutoff can be used to regularized the problem by regularizing $I_{0}(x,E_B)$, which can be written as
\begin{equation}\label{IBregRel}
I_j(x,E_B,\Lambda)=\frac{1}{2\pi }\left( P.V.\int_{-\Lambda }^{\Lambda }\frac{p^{j}e^{ipx}}{E_B-\sqrt{p^{2}+m^{2}}}dp\right).
\end{equation}
We define $I_j( E_B,\Lambda)\equiv I_j(0, E_B,\Lambda)$.
To evaluate $I_0(E_B,\Lambda)$, the right hand side of eq.(\ref{IBregRel}) for $j=0$ is expanded in powers of $E_B/\sqrt{p^2 + m^2}$. This gives
\begin{equation}
I_0(E_B,\Lambda) = - \frac{1}{2\pi} P.V.\int_{-\Lambda} ^{\Lambda} \left(
\frac{1}{\sqrt{p^2 + m^2}}+
\sum_{n=2}^\infty \left(\frac{E_B}{\sqrt{p^2 + m^2}}\right)^n\right)dp.
\end{equation}
By taking the limit $\Lambda\rightarrow \infty$, we find that all the terms in the summation are finite.  On the other hand, the first term is
logarithmically ultra-violet divergent. All the rest of the terms can be integrated separately when $\Lambda\rightarrow \infty$,
and then re-summed. The summation is convergent for a bound state $0<E_B<m$ and a strong bound states $0>E_B>-m$ as it was explained in \cite{Our}. Therefore we get
\begin{equation}
\label{IEB}
\lim _{\Lambda\rightarrow \infty}I_0(E_B,\Lambda)= I_0(E_B)= \lim _{\Lambda\rightarrow \infty}\frac{1}{2\pi }
\log\left(\frac{\sqrt{\Lambda^2+m^2}-\Lambda}{\sqrt{\Lambda^2+m^2}+\Lambda}\right)+I_{0c}(E_B)
\end{equation}
where $I_{0c}(E_B)$ is the finite part of $I_0(E_B,\Lambda)$ as $\Lambda\rightarrow\infty$, in this case it takes the following form
\begin{equation}\label{Ic}
   I_{0c}(E_B)= - \frac{E_B}{2 \pi \sqrt{m^2 - E_B^2}}
\left(\pi + 2 \arcsin\frac{E_B}{m}\right).
\end{equation}
For an ultra-strong bound state with energy $E_B < - m$, the series  diverges as $\Lambda\rightarrow \infty$. Still, the result
can be obtained by directly integrating the convergent expression, and taking the limit $\Lambda\rightarrow \infty$
\begin{eqnarray}
\label{IEBU}
I_{0c}(E_B)&=&\frac{1}{2 \pi} \int dp \ \left(\frac{1}{E_B - \sqrt{p^2 + m^2}} +
\frac{1}{\sqrt{p^2 + m^2}} \right) = \nonumber \\
&&\frac{E_B}{\pi \sqrt{E_B^2 - m^2}} \
\text{arctanh} \left(\frac{\sqrt{E_B^2 - m^2}}{E_B}\right).
\end{eqnarray}
The most elegant expression of $I_{2j}( E_B,\Lambda)$ can be obtained in terms of the hypergeometric function, where the right hand side of eq.(\ref{IBregRel}) can be integrated, and the result is
\begin{eqnarray}\label{IEBhyper}
  I_{2j}(E_B,\Lambda) &=& \frac{1}{\pi}\Lambda^{2j+1}\Bigg(\frac{m F_1(\frac{1}{2}+j;-\frac{1}{2},1; \frac{3}{2}+j; \frac{-\Lambda^2}{m^2},-\frac{\Lambda^2}{m^2-E_B^2})}{(m^2-E_B^2)(2j+1)}\nonumber \\&+&
   \frac{E_B \ _{2}F_1(1,\frac{1}{2}+j; \frac{3}{2}+j; -\frac{\Lambda^2}{m^2-E_B^2})}{(m^2-E_B^2)(2j+1)}\Bigg)(-1)^{j-1},
\end{eqnarray}
where $F_1(a_1;a_2,a_3; a_4;z_1,z_2)$ is the Appel hypergeometric function with two variables, and $_{2}F_1(a_1,a_2; a_3; z)$ is a hypergeometric function with one variable. Interestingly, eq.(\ref{IEBhyper}) is valid for all cases, bound state, strong bound state and ultra-strong bound state. The asymptotic behavior is the same for all these case as $\Lambda\rightarrow \infty$, which is given by the following expression
\begin{equation}\label{IAsymScatRel}
    I_{2j}( E_B,\Lambda)\sim (-1)^{j-1}\frac{\Lambda^{2j}}{2\pi j},  \hspace{20 mm} j=1,2,...
\end{equation}
The expression of the renormalized wave function for this case is
\begin{equation}\label{GeneralPsi2Re}
\Psi (x)=\lim_{\Lambda\rightarrow\infty}\Psi (x,\Lambda)=\lim_{\Lambda\rightarrow\infty}\lambda_n (\Lambda)\sum_{j=0}^{n}C_{j}^{n}
 \Psi_B ^{(n-k)}(0,\Lambda)(-1)^{j}I_j(x,E_B,\Lambda),
\end{equation}
At $x=0$, the above equation leads to $n+1$-system of equations similar to eqs.(\ref{PsizeroSys}) with only $I_j(\Delta E_B,\Lambda)$ replaced by  $I_j(E_B,\Lambda)$. The same argument for deriving the gap equation of the non-relativistic case holds here. Also eq.(\ref{SeqEqEven}), eq.(\ref{SeqEqOdd}) and eq.(\ref{SeqNOE}) are valid for the relativistic case after replacing $I_j(\Delta E_B,\Lambda)$  by  $I_j(E_B,\Lambda)$.

To be a physical state, the wave function for the bound state must be normalizable. From eq.(\ref{FnpEq}) and eq.(\ref{GeneralPsi2Re}). The result is
\begin{eqnarray}\label{NormaRelGen}
  \int_{-\infty}^{\infty}|\Psi_B(x)|^2 dx&=& \lim_{\Lambda\rightarrow \infty}\frac{\lambda_n(\Lambda)^2}{2\pi}\sum_{r=0}^{2n} \sum_{j=0}^{n}i^r C^{n}_{j}C^{n}_{r-j}(-1)^{j}\nonumber\\ &\times&\Psi_B ^{(n-j)}(0,\Lambda)\Psi_B ^{(n-r+j)} (0,\Lambda) \int_{-\Lambda}^{\Lambda}\frac{p^{r}}{(E_B-\sqrt{p^2+m^2})^2} dp
  \nonumber\\&=&1.
\end{eqnarray}
Again here, it seems that the wave function is not normalizable because
\begin{equation}\label{div}
    \int_{-\Lambda}^{\Lambda} \frac{p^{2r}}{(E_B-\sqrt{p^2+m^2})^2} dp\rightarrow\infty,\hspace{20 mm} r= 1,2,...
\end{equation}
On the other hand, further analysis shows that this is not the case, as quantities like $\lambda_n \Psi_B ^{(m)}$  go to zero fast enough as $\Lambda\rightarrow \infty$ for all $m<n$ such that the integrals in eq.(\ref{NormaRelGen}) converges. This will be well demonstrated in section 3.

\begin{figure}[tbh]
\begin{center}
\includegraphics[bb=280 28 400 400,scale=0.45]{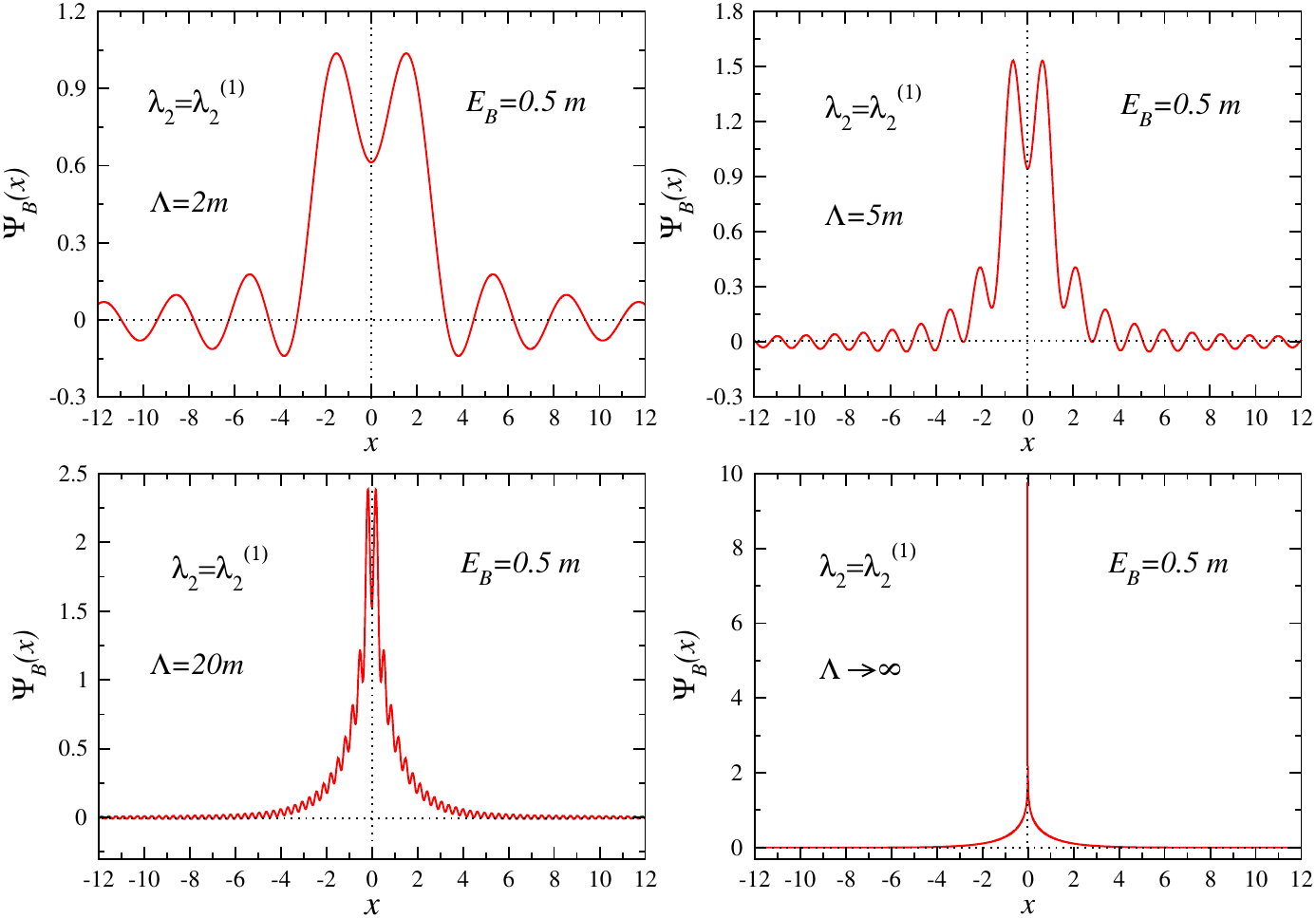}
\vspace{5mm}
\end{center}
\begin{center}
\includegraphics[bb=280 28 400 400,scale=0.45]{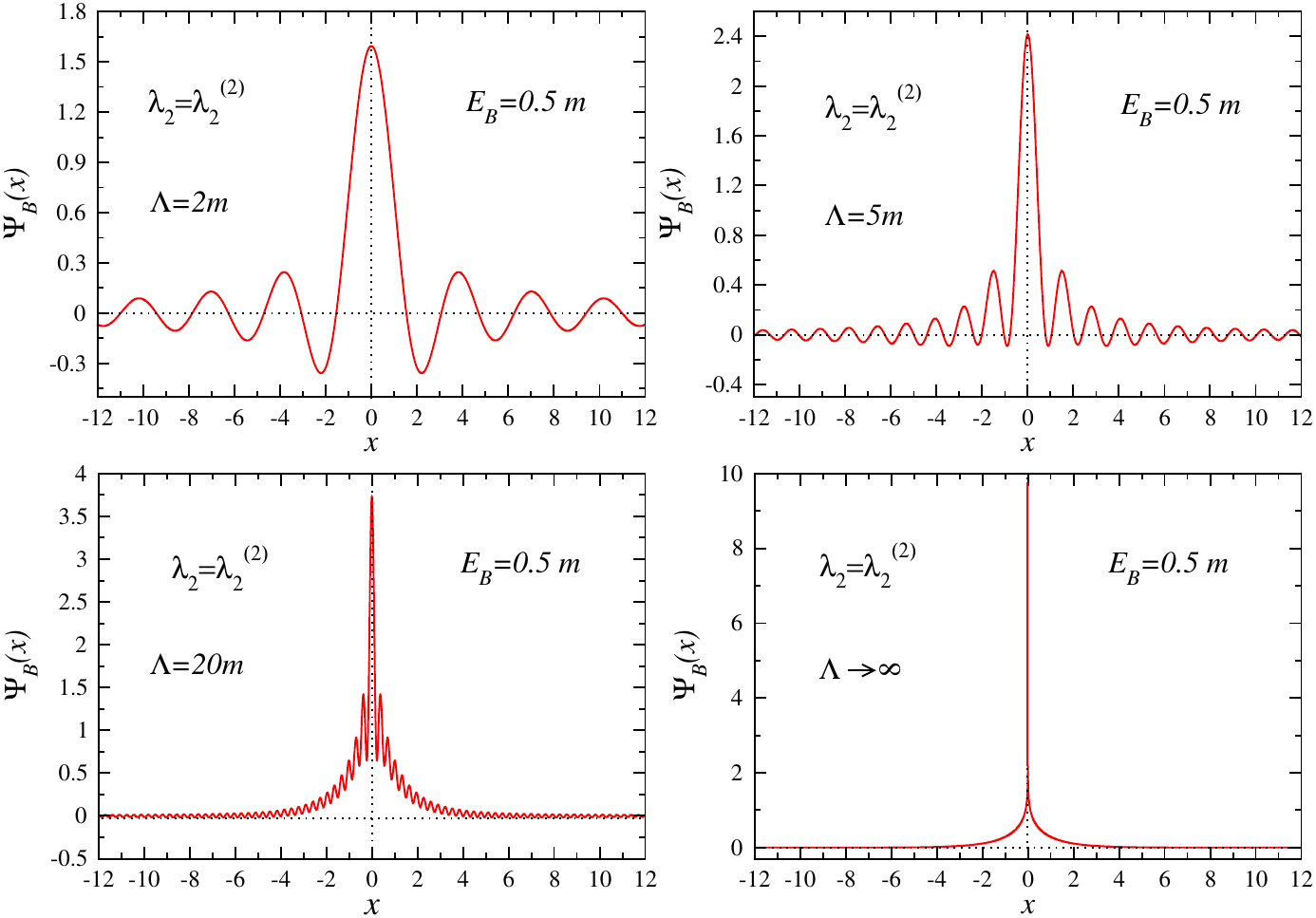}
\vspace{5mm}
\caption{\it Bound state wave function in coordinate space for the relativistic $\delta^{(2)}$-function potential case. The value of the binding energy is $E_B = m/2$, with different values of $\Lambda=2m, 5m,20m$, and $\infty$. For $\lambda_2=\lambda_2^{(1)}$ (top four figures), the curves are different than the analogs ones for the case $\lambda_2=\lambda_2^{(2)}$ (bottom four figures), however, the difference diminish with the increase of $\Lambda$. The wave function in all the figures is an even function for any value of $\Lambda$.}
\end{center}
\label{fig5}
\end{figure}
For the scattering wave function for the relativistic case, we use the following ansatz
\begin{equation}\label{GeneralAnsatzRel}
  \tilde{\Psi}_E(p) =A\delta(p-\sqrt{E^{2}-m^{2}})+B\delta(p+\sqrt{E^{2}-m^{2}})+\tilde{\Phi}_E(p), \hspace{10 mm}  E=\sqrt{k^2+m^2},
\end{equation}
where $A$ and $B$ are arbitrary constants that will be defined later. To calculate the scattering states, we must calculate $\Phi _{E}(x)$. Substituting for $\tilde{\Psi}_E(p)$  from eq.(\ref{GeneralAnsatzRel}) into eq.(\ref{SchNDRel2}), and then solving for   $\tilde{\Phi}_E(p)$ we get
\begin{equation}\label{PhiNonP}
\tilde{\Phi}_E(p)=\frac{\lambda_1
}{E - \sqrt{p^2 + m^2}}\sum_{j=0}^{n} (-1)^{n}(-ip)^{j}C^{n}_{j}\left( \frac{A+(-1)^{n-j}B}{2\pi }(i k)^{n-j}+\Phi^{n-j} _{E}(0)\right).
\end{equation}
In $x$-space,
\begin{equation}\label{GeneralRelXScat}
   \Phi_E(x)=\lambda_n\sum_{j=0}^{n} (-1)^{j+n}C^{n}_{j}I_{2j+1}(x,E)\left( \frac{A+(-1)^{n-j}B}{2\pi }(i k)^{n-j}+\Phi^{n-j} _{E}(0)\right),
\end{equation}
where
\begin{equation}\label{IkRelScat}
I_{j}(x, E)=\frac{1}{2\pi }\left( P.V.\int_{-\infty }^{\infty }\frac{(ip)^{j}e^{ipx}%
}{E-\sqrt{p^2+m^2}}dp\right) =\frac{\partial ^{j}I_{0}(x, E)}{\partial x^{j}%.
}.
\end{equation}
By using momentum cutoff in the expression of $I_{j}(x,E)$ in eq.(\ref{IkRelScat}), can be written in the following form
\begin{equation}\label{IkRelScatReg}
I_{j}(x, E)=\frac{1}{2\pi }\left( P.V.\int_{-\Lambda }^{\Lambda }\frac{(ip)^{j}e^{ipx}%
}{E-\sqrt{p^2+m^2}}dp\right).
\end{equation}
As a result, the function $\Phi_E(x)$ is regularized, where
 \begin{equation}
   \Phi_E(x)=\lim_{\Lambda\rightarrow\infty} \Phi_E(x,\Lambda)
 \end{equation}
This leads to $n+1$ equations that could be obtained from eqs.(\ref{PhiZeroEqs}) by replacing $I_{2j}(\Delta E,\Lambda)$ with $I_{2j}(E,\Lambda)\equiv I_{2j}(0,E,\Lambda)$. Again here
\begin{equation}
    I_{2j+1}( E,\Lambda)=0, \hspace{20 mm}j=0,1,2,...
\end{equation}
The divergent part of $I_0(E,\Lambda)$ is similar to the divergent part of $I_0(E_B,\Lambda)$.
\begin{eqnarray}\label{IE0}
I_0(E,\Lambda) &=& \frac{1}{2\pi }
\log\left(\frac{\sqrt{\Lambda^2+m^2}-\Lambda}{\sqrt{\Lambda^2+m^2}+\Lambda}\right)+ I_{0c}(E)
\end{eqnarray}
where $ I_{0c}(E)$ is the finite part of $I_0(E,\Lambda)$ as $\Lambda\rightarrow\infty$. It can be obtained by using eq.(\ref{IEBU}) for the scattering case, which gives
\begin{equation}\label{I0cE}
    I_{0c}(E)=\frac{E}{\pi \sqrt{E - m^2}} \
\text{arctanh} \left(\frac{\sqrt{E - m^2}}{E}\right).
\end{equation}
Again here   $I_{2j}( E_B,\Lambda)$ can be obtained elegantly in terms of the hypergeometric functions, where the right hand side of eq.(\ref{IkRelScatReg}) can be integrated, and the result is
\begin{eqnarray}\label{IEEhyper}
  I_{2j}(E,\Lambda) &=& \frac{1}{\pi}\Lambda^{2j+1}(-1)^j \Re\Bigg[\frac{m F_1(\frac{1}{2}+j;-\frac{1}{2},1; \frac{3}{2}+j; \frac{-\Lambda^2}{m^2},-\frac{\Lambda^2}{E^2-m^2})}{(E^2-m^2)(2j+1)}\nonumber \\&+&
   \frac{E \ _{2}F_1(1,\frac{1}{2}+j; \frac{3}{2}+j; -\frac{\Lambda^2}{E^2-m^2})}{(E^2-m^2)(2j+1)}\Bigg].
\end{eqnarray}
For $E>m>0$, the functions $F_1(a_1;a_2,a_3; a_4;z_1,z_2)$ and $_{2}F_1(a_1,a_2; a_3; z)$ are multi-value functions. The only relevant expression is the one with  real $I_{2j}(E,\Lambda)$.
The asymptotic behavior of  $I_{2j}(\Delta E,\Lambda)$ as $\Lambda\rightarrow\infty$ is given by the following relation
\begin{equation}\label{IAsymScatRel}
    I_{2j}( E,\Lambda)\sim (-1)^{j-1}\frac{\Lambda^{2j}}{2\pi j},  \hspace{20 mm} j=1,2,...
\end{equation}
\section{The solution for the relativistic $\delta^{(2)}$-potential}
For an even-function potential, the solution is either an even or an odd function. For the even function solution,  eq.(\ref{SeqEqEven}) is applicable, accordingly for $n=2$, the gap equations are
\begin{equation}\label{GapRel2Der}
  \frac{1}{\lambda_2(\Lambda)} =I_{2}(E_B,\Lambda)\pm \sqrt{I_{0}(E_B,\Lambda)I_{4}( E_B,\Lambda)}=\frac{1}{\lambda_2^{(1,2)}(\Lambda)}.
\end{equation}
The regularized form of the wave function for this case is
\begin{equation}\label{RegBoundWF}
  \Psi (x)=\lim_{\Lambda\rightarrow \infty}\frac{\lambda_2 (\Lambda)}{2\pi}\int_{-\Lambda}^{\Lambda}\frac{p^{2}\Psi_B(0,\Lambda)^2-\Psi^{(2)}_B(0,\Lambda)}{E_B-\sqrt{p^2+m^2}} e^{ipx}dp.
\end{equation}
By solving eqs.(\ref{PsizeroSys}), the value of $\Psi_B^{(2)}(0,\Lambda)$ can be obtained in terms of $\Psi_B(0,\Lambda)$. The result for the relativistic case is
\begin{equation}
    \Psi_B^{(2)}(0,\Lambda)=\pm\sqrt{\frac{I_{4}(E_B,\Lambda)}{I_{0}(E_B,\Lambda)}} \Psi_B(0,\Lambda), \hspace{20mm}\lambda_2 (\Lambda)=\lambda_2^{(1,2)}(\Lambda).
\end{equation}
From the asymptotic behavior of $I_{2j}(\Delta E_B,\Lambda)$, we know that $I_{4}(\Delta E_B,\Lambda)\sim \Lambda^{4}$, while $I_{0}(\Delta E_B,\Lambda)$ is logarithmically divergent as $\Lambda\rightarrow\infty$. Therefore, roughly speaking, for large $\Lambda$ in terms of $m$, the value of $\Psi_B^{(2)}(0,\Lambda)$ is a larger quantity than $\Psi_B(0,\Lambda)$.

The wave function must be normalizable. From eq.(\ref{NormaRelGen}), the normalization condition is
\begin{equation}\label{NormSecondRel1}
   \lim_{\Lambda\rightarrow \infty} \frac{\lambda_2(\Lambda)^2}{2\pi}\int_{-\Lambda}^{\Lambda}\frac{p^{4}\Psi_B(0,\Lambda)^2-2p^{2}\Psi_B^{(2)}(0,\Lambda)\Psi_B(0,\Lambda)+\Psi_B^{(2)}(0,\Lambda)^2}{(E_B-\sqrt{p^2+m^2})^2} dp=1.
\end{equation}
In the above expression, and as $\Lambda\rightarrow \infty$, it can be proved that
\begin{eqnarray}\label{DiveQauntities1}
    \int_{-\Lambda}^{\Lambda}\frac{p^4dp}{(E_B-\sqrt{p^2+m^2})^2}&\sim& \Lambda^3, \nonumber\\  \int_{-\Lambda}^{\Lambda}\frac{p^2dp}{(E_B-\sqrt{p^2+m^2})^2}&\sim& \Lambda, \nonumber\\ \int_{-\Lambda}^{\Lambda}\frac{dp}{(E_B-\sqrt{p^2+m^2})^2}&=& \frac{2 E_B}{m^2 - E_B^2} +
\frac{m^2}{(m^2 - E_B^2)^{3/2}}\left(\pi + 2 \arcsin\frac{E_B}{m}\right),
\end{eqnarray}
on the other hand, for this case
\begin{equation}\label{DiveQauntities2}
    \lambda_2(\Lambda)^2 \sim  \frac{1}{\Lambda^4\log\Lambda }, \hspace{20 mm}\Psi_B^{(2)}(0,\Lambda)\sim  \frac{\Lambda^{2}}{\sqrt{\log\Lambda }} \Psi_B(0,\Lambda).
\end{equation}
The above relations mean that as $\Lambda\rightarrow \infty$, the leading term in eq.(\ref{NormSecondRel1}) is the one with $\Psi_B^{(2)}(0,\Lambda)^2$, while all the other terms vanish. For the relativistic case,  we reach here again to the important result that $\lambda_2(\Lambda) \Psi_B^{(2)}(0,\Lambda)$ is a finite non-zero quantity, although $\Psi_B^{(2)}(0,\Lambda)\rightarrow \infty$ as $\Lambda\rightarrow\infty$. The normalization condition gives
\begin{eqnarray}\label{Psi0PPValueF}
  C_1&=& \lim_{\Lambda\rightarrow\infty}\lambda_2(\Lambda)\Psi_B^{(2)}(0,\Lambda)\nonumber\\&=&\pm \sqrt{2\pi}\Bigg(\frac{2 E_B}{m^2 - E_B^2} +
\frac{m^2}{(m^2 - E_B^2)^{3/2}}\left(\pi + 2 \arcsin\frac{E_B}{m}\right)\Bigg)^{-1/2},
\end{eqnarray}
The expression of the bound state wave function can finally be written as
\begin{equation}\label{BRState}
 \Psi_B(x)=C_1\left(I_{0}(x,E_B)\pm \lim_{\Lambda\rightarrow \infty} \sqrt{\frac{I_{0}(E_B,\Lambda)}{I_{4}(E_B,\Lambda)}}I_{2}(x,E_B,\Lambda)\right),
\end{equation}
where $I_{0}(x,E_B)$ is given by eq.(\ref{boundstateI}) for bound and strong bound states, while it is given by eq.(\ref{StboundstateI}) for the ultra-bound state. It can be calculated using contour integral (see Figure 2 bottom panel). As for the second term in the above equation, it can be proved that $I_2(x,E_B,\Lambda)$ has an extremum at $x=0$, where $I_2(E_B,\Lambda)\sim \Lambda^2 $ as $\Lambda\rightarrow \infty$. Nevertheless, the extremum value times $\sqrt{I_{0}(E_B,\Lambda)/I_{4}(E_B,\Lambda)}$ can be neglected relative to the first term which diverges at $x=0$ like $\log\Lambda$ as $\Lambda\rightarrow \infty$.
As in the non-relativistic case, we can not simply say that the second term is zero, because eqs.(\ref{PsizeroSys}) have to be satisfied.

Calculating the scattering states requires calculating $\Phi_E ^{(s)}(0,\Lambda)$. This can be done by using eqs.(\ref{PhiZeroEqs}) for the case $n=2$, which leads to the values of
$\Phi_E (0)$, $\Phi_E ^{(1)}(0)$ and $\Phi_E ^{(2)}(0)$ in terms of $I_{2j}( E_B,\Lambda)$, $I_{2j}( E,\Lambda)$ ($j=0,1,2$), and $ \lambda_2(\Lambda)$. For the even solution, $\Lambda\rightarrow\infty$, the calculations lead to
\begin{eqnarray}\label{Phi0Rel2}
   \lim_{\Lambda\rightarrow\infty}\lambda_2 (\Lambda)\left(-\frac{A+B}{2\pi}k^2+\Phi^{(2)} _{E}(0,\Lambda)\right)&=&\frac{A+B}{2\pi(I_{0}( E_B)-I_{0}( E))},  \nonumber\\ -2\lambda_2(\Lambda) \left(\frac{A-B}{2\pi}ik+\Phi^{(1)} _{E}(0,\Lambda)\right)&\sim&  \frac{1}{\Lambda^2\sqrt{\log\Lambda}}\nonumber\\ \lambda_2(\Lambda) \left(\frac{A+B}{2\pi}+\Phi_{E}(0,\Lambda)\right)&=&\pm\frac{(A+B)\log\Lambda}{2\pi \sqrt{I_4( E_B,\Lambda)}(I_{0}( E_B,\Lambda)-I_{0}( E,\Lambda))} \nonumber\\
   \sim   \frac{\sqrt{\log\Lambda}}{\Lambda^2},  \hspace{10mm}\lambda_2(\Lambda)&=&\lambda_2^{(1,2)}(\Lambda).
\end{eqnarray}
By using eqs.(\ref{PhiZeroEqs}) into eq.(\ref{GeneralNonXScat}), $\Phi_E (x)$ for this case is
\begin{eqnarray}\label{Phid2DerRelxFinal}
 \Phi _{E}(x)&=& (A+B)\Bigg(\frac{1}{4\pi^2 (I_0( E_B)-I_0( E) )} \int_{-\infty }^{\infty }\frac{e^{ipx}dp}{E-\sqrt{p^2+m^2}}\nonumber\\
 &-&\lim_{\Lambda\rightarrow\infty}\frac{1}{\pi}\lambda_2(\Lambda) \left(\frac{A-B}{2\pi}ik+\Phi^{(1)}_{E}(0,\Lambda)\right)\int_{-\Lambda }^{\Lambda}\frac{ip \ e^{ipx}dp}{E-\sqrt{p^2+m^2}}\Bigg),
 \nonumber\\
 &+&\lim_{\Lambda\rightarrow\infty}\frac{1}{2\pi}\lambda_2(\Lambda) \left(\frac{A+B}{2\pi}+\Phi_{E}(0,\Lambda)\right)\int_{-\Lambda }^{\Lambda}\frac{-p^2 \ e^{ipx}dp}{E-\sqrt{p^2+m^2}}\Bigg), \nonumber\\ \lambda_2(\Lambda)&=& \lambda_2^{(1,2)}(\Lambda).
\end{eqnarray}
In the expression of $\Phi _{E}(x)$, the second and third terms vanish as $\Lambda\rightarrow\infty$. In our previous work \cite{Our2}, we have proved that there is a spike in the value of $I_1(x,E,\Lambda)$ in the neighborhood of  $x=\pm \varsigma$. The numerical calculations show that the values of the extrema for $I_{1}(x,E_B,\Lambda)$ is proportional to $\Lambda$, and the value of $ \varsigma$ is inversely proportional to $\Lambda$, as $\Lambda\rightarrow\infty$. This means that the second term in eq.(\ref{Phid2DerRelxFinal}) vanishes for any $x\in(-\infty,\infty)$. For the third term, it is zero except at the point $x=0$, then it is proportional to $\sqrt{\log(\Lambda)}$, however it is still can be ignored in comparison to the first term which diverges as $\log(\Lambda)$. Again here it must be stressed that second and third terms can not be simply put to zero because eqs(\ref{PhiZeroEqs}) must be satisfied. By using eq.(\ref{IEB}), eq.(\ref{IE0}), we find that the expression $I_0(E_B,\Lambda)-I_0(E,\Lambda)=I_{0c}(E_B)-I_{0c}(E)$ is finite, that is because the divergent terms cancel each other. The previous non-relativistic treatment suggests that the energy-dependent relativistic running coupling constant renormalized at the scale $E_B$ is given by following expression \cite{Our}
\begin{eqnarray}\label{lambdaDeri}
    \lambda(E,E_B)=\frac{1}{I_{0 c}(E_B)-I_{0 c}(E)}&=&
- \Bigg [\frac{E_B}{2 \pi \sqrt{m^2 - E_B^2}}
\left(\pi + 2 \arcsin\frac{E_B}{m}\right) \nonumber \\
&+&\frac{E}{\pi \sqrt{E^2 - m^2}}
\text{arctanh}\frac{\sqrt{E^2 - m^2}}{E} \Bigg]^{-1}.
\end{eqnarray}
It is easy to prove that for $\Delta E=E-m\ll m$, and $\Delta E_B= E_B-m \ll -m $, the expression of $\lambda(E,E_B)$ is reduced to the expression of $\lambda(\Delta E_B)$ in eq.(\ref{LambdaEB}). The first integral in the eq.(\ref{Phid2DerRelxFinal}) can be calculated using contour integral (see Figure 4 bottom panel).
From eq.(\ref{Phid2DerRelxFinal}),  and eq.(\ref{GeneralAnsatzRel}), we find that the expression the scattering wave function in $x$-space for this case is
\begin{eqnarray}\label{PsiXdeltaFirstRel}
\Psi_E(x)&=& \Bigg[A e^{ikx} +B e^{-ikx}+
\lambda(E,E_B)(A+B)\sqrt{k^2 + m^2} \frac{\sin(k |x|)}{k}
\nonumber \\
&-&\left. \frac{\lambda(E,E_B)}{\pi} \int_m^\infty d\mu
\frac{\sqrt{\mu^2 - m^2}}{\mu^2 + k^2} \exp(- \mu |x|)\right], \nonumber\\ \lambda_2(\Lambda)&=& \lambda_2^{(1,2)}(\Lambda)
\end{eqnarray}

To understand more the meaning of the wave function in eq.(\ref{PsiXdeltaFirstRel}), and the constants $A$ and $B$,  the reflected and transmitted wave functions for this case must be investigated. In region I to the left of the contact point, i.e. for $x < 0$, the relativistic reflected wave function takes the following form \cite{Our,Our2}
\begin{equation}\label{ReflDel}
\Psi_I(x) = \exp(i k x) + R(k) \exp(- i k x) + C(k) \lambda(E,E_B) \chi_E(x).
\end{equation}
In region II to the right of the contact point, i.e. for $x > 0$, the relativistic transmitted wave function takes the following form
\begin{equation}\label{TransDel}
\Psi_{II}(x) = T(k) \exp(i k x) + C(k) \lambda(E,E_B) \chi_E(x).
\end{equation}
Here, $C(k)$ is a constant that will be determined later, $R(k)$ and $T(k)$ are the reflection and transmission coefficients, and
\begin{equation}
\chi_E(x) = \frac{1}{\pi} \int_m^\infty d\mu \
\frac{\sqrt{\mu^2 - m^2}}{\mu^2 + E^2 - m^2} \exp(- \mu |x|),
\end{equation}
is the branch-cut contribution, which arises in the relativistic case only. This contribution decays exponentially away from the contact point $x = 0$, therefore it has no effect on the scattering wave function at asymptotic distances. By comparing eq.(\ref{PsiXdeltaFirst}) for $x<0$ with eq.(\ref{ReflDel}), and for $x>0$ with eq.(\ref{TransDel}), we get the following relations
\begin{equation}\label{TR}
    T=\frac{k}{k+i\lambda(E,E_B)\sqrt{k^2 + m^2}}, \ \ \ \ \ R=-\frac{i\lambda(E,E_B)\sqrt{k^2 + m^2}}{ k+i\lambda(E,E_B)\sqrt{k^2 + m^2}},
\end{equation}
\begin{eqnarray}
 A=\frac{1}{2}\frac{2k+i\sqrt{k^2 + m^2}}{k+i\sqrt{k^2 + m^2}}, \hspace{10mm} B=\frac{1}{2}R, \hspace{10mm} C(k)=-\frac{1}{\pi}T. \ \ \ \ \ \
\end{eqnarray}
To verify that the resulting system is self-adjoint, we have to prove that the scalar product of the bound state with a scattering state vanishes, or
\begin{equation}\label{ScNonEEB}
    \langle \Psi_B|\Psi_E\rangle=0,\ \ \ \ \ \ \ \lambda_2(\Lambda)= \lambda_2^{(1,2)}(\Lambda),
\end{equation}
we must also prove that the scalar product of a scattering state with energy $E'$ with another scattering state with energy $E$ gives
\begin{equation}\label{ScNonEpE}
\langle \Psi_{E'}|\Psi_E\rangle \sim
\delta(\sqrt{E^2-m^2} - \sqrt{ E'^2-m^2}),\ \ \ \ \ \ \ \ \  \  \lambda_2(\Lambda)= \lambda_2^{(1,2)}(\Lambda).
\end{equation}
Again here, the calculation for proving this are lengthy, however the approached used is similar to the one that is discussed in details in \cite{Our2} appendix B the relativistic part.
\subsection{ Repulsive and Attractive Scattering States, and the Non-relativistic Limit for the Relativistic Case}
For the relativistic case, and once the cutoff is removed, we have the same bound state for both $\lambda_2=\lambda_2^{(1,2)}$, and the same scattering states for both $\lambda_2=\lambda_2^{(1,2)}$. Moreover, the wave function of the scattering state is similar to the one for the $\delta$-function potential and the $\delta'$-function potential. To elucidate that, consider the even part of the wave function in eq.(\ref{PsiXdeltaFirstRel})
\begin{eqnarray}
\frac{\Psi_E(x)+\Psi_E(-x)}{2}&=&A' \left[\cos(k x) +
\lambda(E,E_B) \frac{\sqrt{k^2 + m^2}}{k} \sin(k |x|) \right.
\nonumber \\
&-&\left. \frac{\lambda(E,E_B)}{\pi} \int_m^\infty d\mu
\frac{\sqrt{\mu^2 - m^2}}{\mu^2 + k^2} \exp(- \mu |x|)\right],
\end{eqnarray}
where $A'=A+B$. This exactly the same expression of the scattering wave function of the $\delta$-function and $\delta'$- potentials that was derived in \cite{Our,Our2}. The same goes for the bound state.

From eq.(\ref{lambdaDeri}) and eq.(\ref{Ic}), bound and strong bound states ($|E_B|<m$) are correspond to attractive  $\delta^{(2)}$-function potential, because then $\lambda(E,E_B)<0$ . On the other hand, for ultra- strong bound state ($ E_B < - m $ ), the value of $I_{0c}(E_B)$ is given by eq.(\ref{IEBU}), and therefore it gives $ \lambda(E,E_B)>0$ for $E>E_B$, (see Figure 6). This correspond to  a repulsive $ \delta^{(2)}$-function potential. The non-relativistic cutoff regularization for the even solution of the $ \delta^{(2)}$-function potential can not lead to a repulsive solution, but only to an attractive one, as it was explained in section 3.
\begin{figure}[tbh]
\begin{center}
\includegraphics[bb=280 28 400 400,scale=0.6]{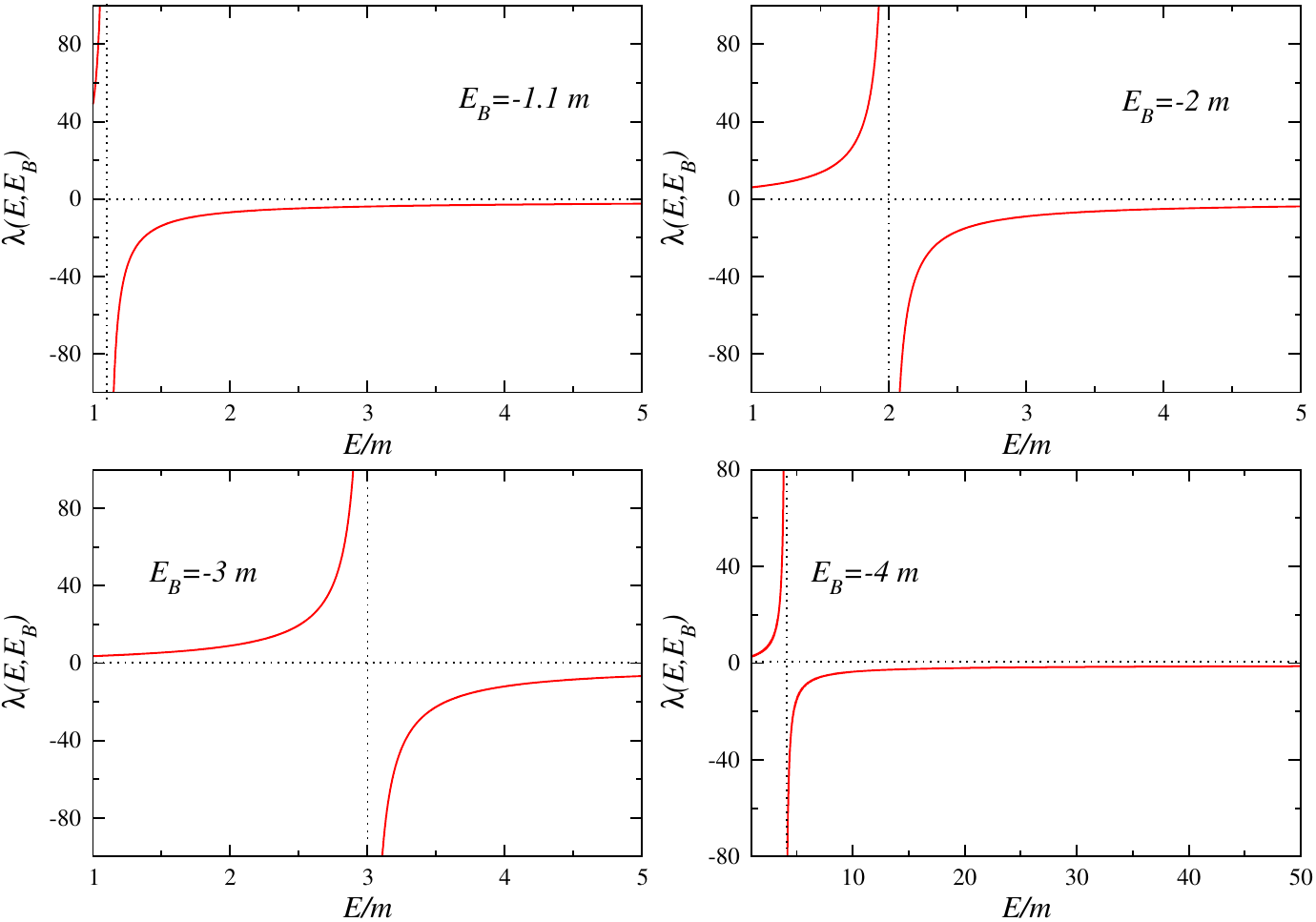}
\end{center}
\vspace{5mm}
\caption{\it The running coupling $\lambda(E,E_B)$ as a function of the scattering energy $E$ in the units of $m$, for $E_B=-1.1m, -2m, -3m,$ and $-4m$. The graph in the lower right corner was extended to large values of $E$ in order to illustrate the asymptotic freedom of the system when $\lambda(E,E_B)\rightarrow 0$ as $E\rightarrow \infty$}
\label{fig6}
\end{figure}
By taking $\varkappa/m \rightarrow 0$, we get the non-relativistic limit for the relativistic bound state. Accordingly, eq.(\ref{BRState}) gives
\begin{eqnarray}\label{NRbou}
\Psi_B(x) &=&\sqrt{\varkappa}
\Big[\frac{\varkappa}{m \pi}  \int_m^\infty d\mu
\frac{\sqrt{\mu^2 - m^2}}{\mu^2 - \varkappa^2} \exp(- \mu |x|) +
\exp(- \varkappa |x|)\Big].
\end{eqnarray}
This means that the wave function
reduces to the bound state for the non-relativistic case in eq.(\ref{NonRelWFfinal1}). However, the divergence at the origin of the relativistic wave function persists for any
non-zero value of $\varkappa/m$. The non-relativistic limit for the relativistic scattering state is
\begin{eqnarray}\label{PsiXdeltaFirstNonLim}
\Psi_E(x)&=& A e^{ikx} +B e^{-ikx}+
(A+B)\lambda(E_B) \frac{m \sin(k |x|)}{k} \nonumber \\
&-&\left(\frac{1}{\pi}\lambda(E_B)(A+B)\int_m^\infty d\mu
\frac{\sqrt{\mu^2 - m^2}}{\mu^2 + k^2} \exp(- \mu |x|)\right), \nonumber \\ \hspace{10mm}\lambda_2(\Lambda)&=&\lambda_2^{(1,2)}(\Lambda),
\end{eqnarray}
where $E=k^2/ 2m$. Again here, the divergence at the origin of the relativistic wave function persists.
\par By taking the limit $E_B \rightarrow - \infty$, the running coupling constant in eq.(\ref{lambdaDeri}) can be written as
\begin{eqnarray}
\lambda(E,E_B) &\rightarrow& - \left[\frac{E}{\pi \sqrt{E^2 - m^2}} \
\text{arctanh}\frac{\sqrt{E^2 - m^2}}{E} -
\frac{1}{\pi} \log\left(\frac{- 2 E_B}{m}\right)\right]^{-1},  \nonumber \\ \hspace{10mm}\lambda_2(\Lambda)&=&\lambda_2^{(1,2)}(\Lambda).
\end{eqnarray}
For small non-relativistic energies $\Delta E = E - m \ll m$, this reduces to
\begin{equation}\label{LambdaEBg}
\lambda \rightarrow \frac{\pi}{\log(- 2 E_B/m)} > 0.
\end{equation}
Therefore we are reaching the non-relativistic limit for a repulsive $\delta^{(2)}$-function potential with a coupling parameter $\lambda(E_B)>0$. This exactly the same as the case of the $\delta$- function and $\delta'$ potentials that were discussed in \cite{Our,Our2}. An important feature of the non-relativistic case is that it has only an attractive $\delta^{(2)}$-function potential. In contrast, the non-relativistic limit of the relativistic case for  ultra-strong bound state gives a repulsive $\delta^{(2)}$-function potential with $\lambda(E_B)>0$ in eq.(\ref{PsiXdeltaFirstNonLim}). At first glance, this seems to be a paradox. However, the fact that contact interactions happen at very short distances can explain the issue. Very short distances mean high momentum transfer, therefore even for non-relativistic limit energies, the particle still influenced by the powers of $p$ higher than two in the expansion of the pseudo-differential operator.
\subsection{The odd wave function solution}
For the odd function solution, $\Psi_B(0,\Lambda)=\Psi_B^{(2)}(0,\Lambda)=0$, for this case  eq.(\ref{SeqEqOdd}) is applicable. Accordingly for $n=2$, the gap equation is
\begin{equation}\label{Gap2DerRelOdd}
  \frac{1}{\lambda_2(\Lambda)} =-2I_2( E_B,\Lambda).
\end{equation}
The regularized form of the wave function for this case is
\begin{equation}\label{RegBoundWF}
  \Psi_B(x)=\frac{1}{2\pi}\lim_{\Lambda\rightarrow \infty}\lambda_2 (\Lambda)\int_{-\Lambda}^{\Lambda}\frac{-2ip\Psi_B^{(1)}(0,\Lambda)}{E_B-\sqrt{p^2+m^2}}e^{ipx}dp.
\end{equation}
The wave function must be normalizable. From eq.(\ref{NormSecondRel1}), the normalization condition is
\begin{equation}\label{NormSecond1}
   \lim_{\Lambda\rightarrow \infty} \frac{\lambda_2(\Lambda)^2}{2\pi}\int_{-\Lambda}^{\Lambda}\frac{4p^{2}\Psi_B^{(1)}(0,\Lambda)^2}{(E_B-\sqrt{p^2+m^2})^2} dp=1.
\end{equation}
In the above expression, as $\Lambda\rightarrow \infty$
\begin{equation}\label{DiveQauntities1}
\int_{-\Lambda}^{\Lambda}\frac{p^2}{(E_B-\sqrt{p^2+m^2})^2}dp\sim 2\Lambda,
\end{equation}
also, for this case, and  as $\Lambda\rightarrow \infty$
\begin{equation}\label{DiveQauntities2}
    \lambda_2(\Lambda)^2 \sim \frac{1}{\Lambda^4}.
\end{equation}
As a result, the normalization condition gives
\begin{equation}
   \lambda_2(\Lambda)\Psi_B^{(1)}(0,\Lambda)\sim \frac{1}{\sqrt{8\Lambda}},
\end{equation}
and
\begin{equation}
   \Psi_B^{(1)}(0,\Lambda)\sim \Lambda^{3/2}.
\end{equation}
Therefor we can write eq.(\ref{RegBoundWF}) as
\begin{equation}\label{RegBoundWF}
  \Psi_B(x)=\frac{1}{2\pi}\lim_{\Lambda\rightarrow \infty}\frac{1}{\sqrt{2\Lambda}}\int_{-\Lambda}^{\Lambda}\frac{-ip}{E_B-\sqrt{p^2+m^2}}e^{ipx}dp.
\end{equation}
This is a normalizable wave function, but it is highly localized because of the factor $1/\sqrt{\Lambda}$.

For the scattering states, the expressions of $\Phi_E (0,\Lambda)$, $\Phi_E ^{(1)}(0,\Lambda)$ and $\Phi_E ^{(2)}(0,\Lambda)$ can be calculated this time too using $\lambda_2(\Lambda)$ from eq.(\ref{Gap2DerRelOdd}). As $\Lambda\rightarrow\infty$, the calculations lead to
\begin{eqnarray}\label{Phi0RelOdd2}
   \lambda_2 (\Lambda)\left(-\frac{A+B}{2\pi}k^2+\Phi^{(2)} _{E}(0,\Lambda)\right)&\sim &-\frac{A+B}{2\pi I_{0}( E,\Lambda)}\sim \frac{1}{\log \Lambda},  \nonumber\\ -2\lambda_2(\Lambda) \left(\frac{A-B}{2\pi}ik+\Phi^{(1)} _{E}(0,\Lambda)\right)&=&-ik\frac{A-B}{2\pi (I_2( E_B,\Lambda)-I_2( E,\Lambda))}\sim \frac{1}{\Lambda}\nonumber\\  \lambda_2(\Lambda)\left(\frac{A+B}{2\pi}+\Phi_{E}(0,\Lambda)\right)&\sim & \frac{1}{\Lambda^2 \log \Lambda}
\end{eqnarray}
From the above equations, and as $ \Lambda\rightarrow\infty$, the scattering wave function for this case is
\begin{eqnarray}\label{RelPsiScatOdd}
 \Psi _{E}(x)&=& A e^{ikx} +B e^{-ikx}+ \lim_{\Lambda\rightarrow\infty}\frac{A+B}{2\log\Lambda} \int_{-\infty }^{\infty }\frac{e^{ipx}dp}{E-\sqrt{p^2+m^2}}\nonumber\\
 &+&\lim_{\Lambda\rightarrow\infty}\frac{ik(A-B)}{E_B-E}\frac{1}{\Lambda}\int_{-\infty }^{\infty }\frac{ipe^{ipx}dp}{E-\sqrt{p^2+m^2}}.
\end{eqnarray}
In region I to the left of the contact point, i.e. for $x < 0$, the relativistic reflected wave function is
\begin{eqnarray}\label{ReflRelOdd}
\Psi_I(x) &=& \exp(i k x) + R(k) \exp(- i k x) +\lim_{\Lambda\rightarrow\infty} \frac{A+B}{2\log\Lambda}\int_{-\Lambda }^{\Lambda }\frac{e^{ipx}dp}{E-\sqrt{p^2+m^2}}\nonumber\\&+&\lim_{\Lambda\rightarrow\infty}\frac{ik(A-B)}{E_B-E}\frac{1}{\Lambda}\int_{-\Lambda }^{\Lambda}\frac{ipe^{ipx}dp}{E-\sqrt{p^2+m^2}}
\end{eqnarray}
In region II to the right of the contact point, i.e. for $x > 0$, the relativistic transmitted wave function takes the following form
\begin{eqnarray}\label{TransRelOdd}
\Psi_{II}(x) &=& T(k) \exp(i k x) + \lim_{\Lambda\rightarrow\infty}\frac{A+B}{2\log\Lambda}\int_{-\Lambda }^{\Lambda }\frac{e^{ipx}dp}{E-\sqrt{p^2+m^2}}\nonumber\\&+&\lim_{\Lambda\rightarrow\infty}\frac{ik(A-B)}{E_B-E}\frac{1}{\Lambda}\int_{-\Lambda}^{\Lambda}\frac{ipe^{ipx}dp}{E-\sqrt{p^2+m^2}}.
\end{eqnarray}
By comparing eq.(\ref{ReflRelOdd}), and eq.(\ref{TransRelOdd}) with eq.(\ref{RelPsiScatOdd}), the only possible solution is
\begin{equation}\label{TRodd}
   B= R(k)=0, \hspace{20mm} A=T(k)=1.
\end{equation}
However there is a delicate properties of the scattering wave function at the neighborhood of the contact point. In the appendix, it has been shown that as $\Lambda\rightarrow\infty$, the second integral in eq.(\ref{RelPsiScatOdd}) has exterma at $x=\pm a_m(E)\Lambda^{-1}$, where the values of $a_1(E),a_2(E),...$ can be evaluated only numerically at this stage. As it is shown in Figure 7 $a_1(E)<a_2(E)<a_3(E)....$. The values of the second integral at these points is $\pm b_m(E)\Lambda$. Again here, $b_1(E),b_2(E),...$ can be evaluated only numerically. In Figure 7 we find that $b_1(E)>b_2(E)>b_3(E)>....$. As a result of the previous discussion
\begin{equation}\label{PsiOriginII}
    \Psi_{II}(-a_m(E)\Lambda^{-1})=-b_m(E)\frac{ik(A-B)}{E_B-E}, \hspace{20mm} m=1,2,...,
\end{equation}
and
\begin{equation}\label{PsiOriginI}
    \Psi_I(a_m(E)\Lambda^{-1})=b_m(E)\frac{ik(A-B)}{E_B-E}, \hspace{20mm} m=1,2,...
\end{equation}
The properties in eq.(\ref{PsiOriginII}) and eq.(\ref{PsiOriginI}) does not influence the behavior of the scattering wave function. In fact, once the cutoff is removed the value of $T=1$, and  $R=0$, which means that the resulting scattering states  $\Psi_E(x)$ are  the ones for a free particle. As a result of the argument in this subsection, the odd wave function solution has a bound state which is normalizable, on the other hand, the particle does not scatter from the potential, instead it acts as the potential does not exist.
\section{Summary and Conclusions}
A general method has been developed to solve the Schr\"odinger equation relativistically and non-relativistically for an arbitrary derivative of the $\delta$ -function potential in 1-d. The problem needed to be regularized in both of the two cases. The method of choices and convenience is cutoff regularization. As we know, when the $n$-derivative of the delta function potential is an even number, the bound state solution is either an even or an odd function. On the other hand when $n$ is an odd number. A separated procedures has been developed to deal with the even $n$ case and with the odd $n$ case, which is valid relativistically and non-relativistically. It has been proved that the even $n$-derivative leads to $2n-1$ gap equations $n$ of  associated with the even-function solution, while $n-1$ gap equations associated with the odd-function solution. The odd $n$-derivative case leads to $n+1$ gap equations. In each equation, the bare $\lambda_n$ coupling constant is expressed in term of the integral $I_{2j}(\Delta E_B,\Lambda)$ for the non-relativistic case given be eq.(\ref{IkHyper}), or $I_{2j}( E_B,\Lambda)$ in the relativistic case given be eq.(\ref{IEBhyper}).  The value of any $\Psi^{(s)}(0,\Lambda)$ with $s=0,...n-1$ can be expressed in terms of $\Psi^{(n)}(0,\Lambda)$  using eq.(\ref{PsizeroSys}). The treatment leads to $\lambda_n(\Lambda)\rightarrow 0$ as $\Lambda\rightarrow\infty$, on the other hand $\lambda_n(\Lambda)$ always appears in all the formulation as $\lambda_n (\Lambda)\Psi_B^{(s)}(0,\Lambda)$, where $s=0,1,...n$. If the problem is renormailzable, then $\lim_{\Lambda\rightarrow\infty}\lambda_n (\Lambda)\Psi_B^{(s)}(0,\Lambda)=0$ for $s=0,...n-1$, while $\lim_{\Lambda\rightarrow\infty}\lambda_n (\Lambda)\Psi_B^{(n)}(0,\Lambda)=C_1$, and the value of $C_1$ is dictated by the normalization condition. Under the same regularization scheme, the scattering wave function can be derived relativistically and non-relativistically for arbitrary $n$.

As an application for the method, the  $\delta^{(2)}$- function potential has been used as an example. For the even solution case $\Psi'_B(0)=0$. It has been shown that there are two possible values for the bare coupling constant $\lambda^{(1,2)}(\Lambda)$ given in  eq.(\ref{Gap2Der}), and eq.(\ref{GapRel2Der}). This example is highlighting the importance of redefining the concept of renormalization from renormalizing $\lambda^{(1,2)}$ to renormalizing the combination $\lambda^{(1,2)} \Psi_B(0)$, which vanishes as $\Lambda\rightarrow \infty$, and renormalizing   $\lim_{\Lambda\rightarrow\infty}\lambda_2 \Psi_B^{(2)}(0)$ which is equal to the normalization constant $\pm C_1$. For the non-relativistic odd solution case, the problem is proved to be non-renormalizable. That is because there is an odd-function bound state given by eq.(\ref{NonBoundOdd}), but there is no scattering state because $\Phi_E(x)$ diverges for any value of $x$ when the cutoff is removed. For the relativistic odd solution case, there is a normalizable bound state, but it is highly local, which means that it is only nonzero values in the neighborhood of $x=0$. Moreover, the scattering solution leads only to a trivial free particle solution when the cutoff is removed. This is analogous to some quantum filed theories, which reduce to a trivial free solution once the cutoff is removed.

Another result of this work is in highlighting the fact that the non-relativistic limit of the relativistic case does not lead exactly to the non-relativistic solution. That is because the  non-relativistic case  has only an attractive $\delta^{(2)}$-function potential. In contrast, the non-relativistic limit of the relativistic case, and for ultra-strong bound state gives a repulsive $\delta^{(2)}$-function potential, where $\lambda(E_B)>0$. This is explained by the fact  that $\delta^{(2)}$-function potential is a contact interaction that takes place at very short distances, which mean high momentum transfer is taking place. Therefore, even for non-relativistic energies limit, the particle still influenced by powers of $p$ higher than two in the expansion of the pseudo-differential operator. This  also explains why the divergence at the origin persist when taking the non-relativistic limit of the relativistic case. Both of the $\delta^{(2)}$-function and $\delta'$-function potentials reveal this feature more than the $\delta$-function potential, because in the the $\delta$-function potential there is a repulsive solution for the non-relativistic case.

For the relativistic and non-relativistic $\delta^{(2)}$-function potential even solution, there are 2-parameters family of self-adjoint extension parameters $\lambda_2(\Lambda)=\lambda^{(1,2)}$. When $\Lambda$ is not large, the wave function for the bound state is an even function, which is differ than the $\delta$-function potential bound state wave function. The wave function for $\lambda_2=\lambda^{(1)}$ is different from the the one for $\lambda_2=\lambda^{(2)}$. As the momentum cutoff value increases, the wave function becomes increasingly similar, and the difference between the the two case diminish. When the cutoff is removed, we get an even wave function with no difference what so ever between $\lambda_2= \lambda^{(1)}$ and the $\lambda_2= \lambda^{(2)}$, as one can see this from Figure 3  and Figure 5. This means that when removing the cutoff, only one parameter left, that is the coupling constant $\lambda(\Delta E_B)$ in the non-relativistic case, and  $\lambda(E,E_B)$ energy-dependent relativistic running coupling constant in the relativistic case. This is similar to the situation in local quantum field theories, when all the Lagrangians correspond to different models reduce to a one Lagrangian once the cutoff is removed. Then, we left only with certain terms and their associated parameters, while all the other parameters associated with the vanishing term become irrelevant. It is important to note here that the non-relativistic self-adjoint extension theory predict also that there is only one parameter. After removing the cutoff, relativistic even solution leads to wave functions for bound and scattering states with the same expression of the analogous ones in the relativistic $\delta$-function and $\delta'$-function potentials. As a result, we have the same interesting features like, asymptotic freedom, dimensional transmutation, and an infra-red conformal fixed point in the massless limit that was discussed in our previous paper \cite{Our}. Moreover, the calculations show that the bound state of the potential $\lambda_2\delta^{(2)}(x)+\lambda_1\delta^{(1)}(x)+\lambda \delta(x)$ has exactly the same expression of wave function for the bound state of the $\delta$-function potential when the cutoff is removed. Also with one parameter survive the cutoff regularization. It is not clear if the work in \cite{Alb97} leads to the same conclusion.

The previous results show that the evidence of universality is extended as far as the relativistic $\delta^{(i)}$-function potentials are concerned, where $i=0,1,2$. That is because in each of these problems, all the parameters are reduced to only one parameter once the cutoff is removed. In addition the $\delta^{(i)}$-function potentials have the same wave functions after removing the cutoff. This means that the particle in question is blind to the difference between the $\delta^{(2)}$-function,  $\delta^{(1)}$-function, and the $\delta$-function potentials, or a combination of the three of them. This is  highly a non-trivial result because it means that there is an  additional feature that relativistic quantum mechanics shares with local quantum field theories, that is universality. This does not mean that the Hamiltonian for the $\delta^{(2)}$-function potential is local, in fact it isn't because of the nonlocal operator $\sqrt{p^2+m^2}$. Does universality hold for a more general contact interaction like $\sum_{n=0}^{k}\lambda_n \delta^{(n)}$ is still an open question, even after this study. To prove that there is universality to all relativistic contact interactions, we have to prove that the outcome is insensitive to any details of the interaction after removing the cutoff, and the result is always similar to the $\delta$-function potential case. The non-relativistic $\delta^{(2)}$-function potential  Hamiltonian is local, and when removing the cutoff, only one parameter remains. Nevertheless, we can not say that there is an evidence of universality, because the the behavior in this case is different than the non-relativistic $\delta$-function potential case which has a repulsive solution, while  $\delta^{(2)}$-function potential has only attractive solution.

In our opinion, the most important outcome of this study is demonstrating an exercise in applied mathematics, on how to quantify singularities for certain singular functions in term of the cutoff value, or, in layman's terms, how small is small, and how big is big in terms of $\Lambda$. The best part of this demonstration is in describing the behavior of the scattering wave function in the neighborhood of $x=0$ for the odd case. This is done by combining analytic and numerical methods in calculating $I_1(x,E,\Lambda)$, which is one of the terms in the expression of $\Psi_E(x)$. In the appendix it was shown that there is a certain type of behavior of $I_1(x,E,\Lambda)$ in neighborhood of $x=0$, which does not change as $\Lambda$ pushed further and further to infinity, as Figure 7 illustrate.
\section*{Acknowledgments}
This publication was made possible by the NPRP grant \# NPRP 5 - 261-1-054 from
the Qatar National Research Fund (a member of the Qatar Foundation). The
statements made herein are solely the responsibility of the authors.
\section*{Appendix : The behavior of $I_1(x,E,\Lambda)$ in the neighborhood of the contact point}
The expression of $I_1(x,E,\Lambda)$ can be obtained from eq.(\ref{IkRelScatReg}). It takes the following form
\begin{eqnarray}
I_1(x,E,\Lambda)=\frac{1}{2\pi}P.V.\left(\int_{-\Lambda}^{\Lambda}\frac{i p \exp(ip x)}{E-\sqrt{p^2+m^2}} dp \right)
\end{eqnarray}
In the neighborhood of the contact point the exponential can be expanded, and therefore the above equation can be written as
 \begin{eqnarray}
I_1(x,E,\Lambda)=\frac{1}{2\pi}P.V.\left(\sum_{j=1}^{\infty} \int_{-\Lambda}^{\Lambda}\frac{(i p)^j x^{j-1}}{(E-\sqrt{p^2+m^2})(j-1)!} dp. \right)
\end{eqnarray}
The above relation can be written in terms of $I_{2j}(E,\Lambda)$. The result is
\begin{equation}
  I_1(x,E,\Lambda)= \sum_{j=1}^{\infty} \frac{I_{2j}(E,\Lambda)x^{2j-1}}{(2j-1)!}.
\end{equation}
Also
\begin{eqnarray}
I_0(x,E,\Lambda)=\frac{1}{2\pi}P.V.\left(\sum_{j=0}^{\infty} \int_{-\Lambda}^{\Lambda}\frac{(i p)^j x}{(E-\sqrt{p^2+m^2})j!} dp. \right),
\end{eqnarray}
or
\begin{equation}
  I_0(x,E,\Lambda)= \sum_{j=0}^{\infty} \frac{I_{2j}(E,\Lambda)x^{2j}}{(2j)!}.
\end{equation}
\\
\\
\begin{figure}[H]
\begin{center}
\includegraphics[bb=280 28 400 400,scale=0.6]{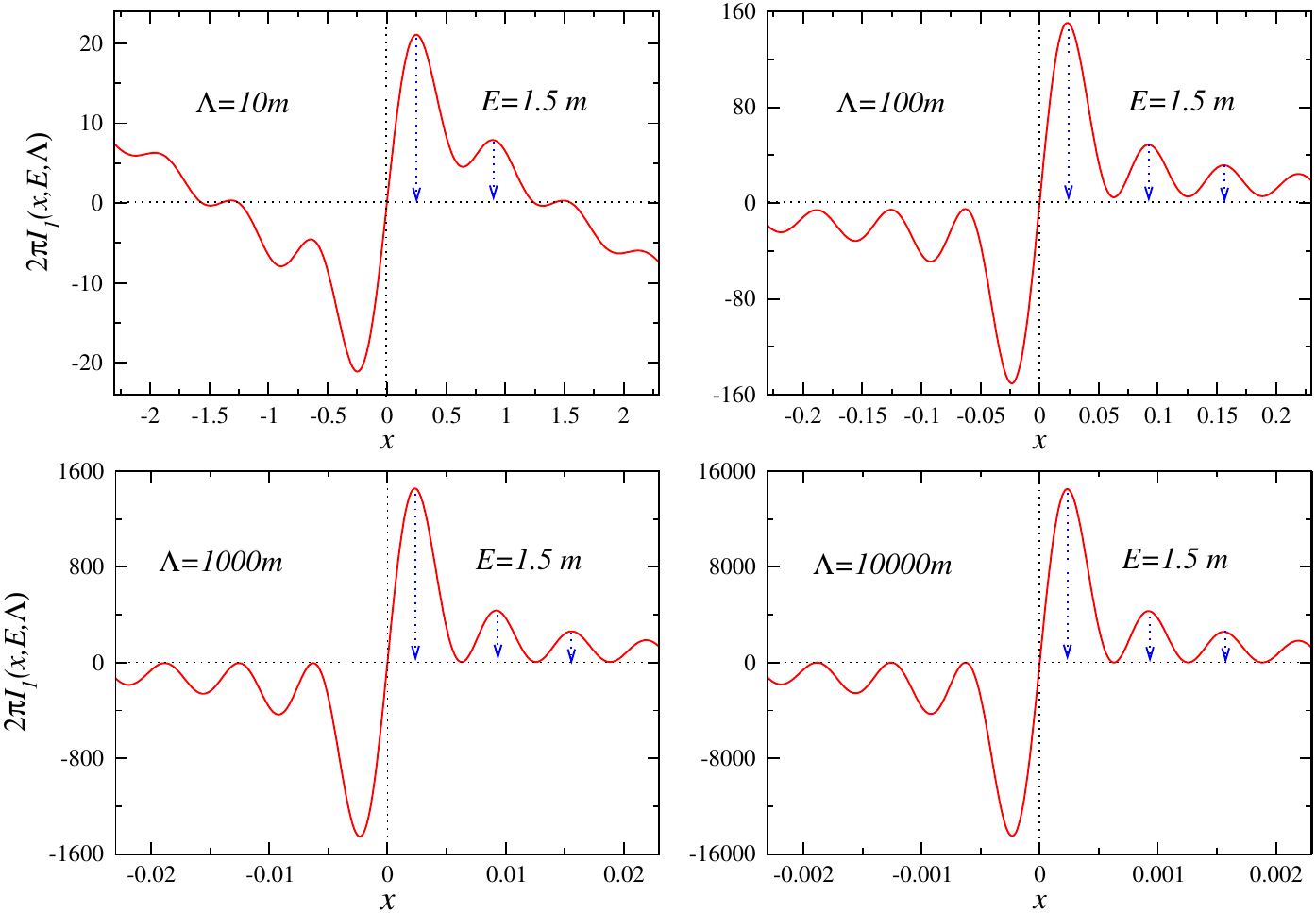}
\vspace{5mm}
\caption{\it The plot of $2\pi I_0(x,E,\Lambda)$ versus $x$ in the neighborhood of $x=0$, for $\Lambda=10 m, 10^{2}m,10^{3}m$ and $10^{4}m$. The exterma at $x=\pm a_m(E)\Lambda^{-1}$, where the values of $a_1(E),a_2(E),...$ can be evaluated numerically. Here, $a_1(E)<a_2(E)<a_3(E)....$. The exterma values are $\pm b_m(E)\Lambda$. Again here, $b_1(E),b_2(E),...$ can be evaluated numerically, where  $b_1(E)>b_2(E)>b_3(E)>....$. B.}
\label{fig7}
\end{center}
\end{figure}

\end{document}